\pdfoutput=1
\documentclass[letterpaper,titlepage,11pt]{article}

\usepackage{amssymb,amsmath,amsfonts,jheppub}
\usepackage{mathrsfs}
\usepackage{epsfig}
\usepackage[toc]{appendix}
\numberwithin{equation}{section}
\usepackage{graphicx}
\usepackage{subfig}
\usepackage{tikz}
\usepackage{tensor}
\usepackage{todonotes}
\usetikzlibrary{decorations.pathmorphing,calc}

\usepackage[normalem]{ulem}

\usepackage{amsmath,amssymb}
\usepackage{verbatim}
\usepackage{graphicx}
\usepackage{color}

\DeclareFontFamily{OT1}{rsfs}{}
\DeclareFontShape{OT1}{rsfs}{m}{n}{ <-7> rsfs5 <7-10> rsfs7 <10->rsfs10}{}
\DeclareMathAlphabet{\mycal}{OT1}{rsfs}{m}{n}

\newcommand{\unity}{1\hspace{-0.243em}\text{l}}

\newcommand{\be}[1]{ \begin{equation}\label{#1}
}

\newcommand{\bea}[1]{\begin{eqnarray}\label{#1}
}
\newcommand{\eea}{
\end{eqnarray}}

\newcommand{\tr}{\textrm{tr}}

\newcommand{\eq}[2]{\begin{equation} #1 \label{#2} \end{equation}}
\newcommand{\eps}{\varepsilon}

\DeclareMathOperator{\extdm}{d}
\newcommand{\extd}{\extdm \!}

\newcommand{\vp}{\varphi}
\newcommand{\ve}{\varepsilon}
\newcommand{\cL}{\mathcal{L}}

\def\cA{\mathcal{A}}
\def\cB{\mathcal{B}}
\def\cC{\mathcal{C}}
\def\cD{\mathcal{D}}
\def\cE{\mathcal{E}}
\def\cF{\mathcal{F}}
\def\cG{\mathcal{G}}
\def\cH{\mathcal{H}}

\def\cK{\mathcal{K}}
\def\cL{\mathcal{L}}

\def\cO{\mathcal{O}}
\def\cP{\mathcal{P}}
\def\cQ{\mathcal{Q}}
\def\cR{\mathcal{R}}

\def\cW{\mathcal{W}}
\def\cX{\mathcal{X}}

\def\cZ{\mathcal{Z}}

\newcommand{\dd}{\mathrm{d}}
\renewcommand{\d}{\partial}

\newcommand{\mytitle}{Towards a bulk description of higher spin SYK}

\title{\mytitle}

\subheader{TUW--18--02}

\author{Hern\'an A. Gonz\'alez,}
\emailAdd{hgonzale@}
\author{Daniel Grumiller,}
\emailAdd{grumil@}
\author{and Jakob Salzer}
\emailAdd{salzer@hep.itp.tuwien.ac.at}
\affiliation{Institute for Theoretical Physics, TU Wien, Wiedner Hauptstr.~8-10/136, A-1040 Vienna, Austria}

\abstract{
We consider on the bulk side extensions of the Sachdev--Ye--Kitaev (SYK) model to Yang--Mills and higher spins. To this end we study generalizations of the Jackiw--Teitelboim (JT) model in the BF formulation. Our main goal is to obtain generalizations of the Schwarzian action, which we achieve in two ways: by considering the on-shell action supplemented by suitable boundary terms compatible with all symmetries, and by applying the Lee--Wald--Zoupas formalism to analyze the symplectic structure of dilaton gravity. We conclude with a discussion of the entropy (including log-corrections from higher spins) and a holographic dictionary for the generalized SYK/JT correspondence.
}


\begin{document}

\maketitle

\section{Introduction}\label{se:1}

The Sachdev--Ye--Kitaev (SYK) model \cite{Kitaev:15ur, Sachdev:1992fk, Sachdev:2010um, Maldacena:2016hyu} describes a quantum mechanical system that holographically produces the behavior of dilaton gravity in two dimensions. It consists of $N$ Majorana fermions at finite temperature $T=\beta^{-1}$ interacting with each other through 4-Fermi interactions with random couplings characterized by the random coupling strength $J$. At low temperatures, $T\ll J$, in the large $N$ limit, $N\gg 1$, the system develops conformal symmetry in one dimension, which is spontaneously broken due to finite temperature effects. 

In this limit the SYK model is effectively controlled by a field $\phi(\tau)$ whose dynamics is governed by the Schwarzian action
\eq{
I[\phi(\tau)]=\frac{N}{\beta J}\,\int\limits^\beta_0  \extd\tau\, \big[\tfrac{1}{2}\,\phi'^2+\{\phi \, ;\tau\} \big]
}{eq:syk1}
where
\eq{
\{\phi \, ;\tau\}=\frac{{\phi'''}}{\phi'}-\frac{3}{2}\,\left(\frac{\phi''}{\phi'}\right)^2
}{eq:syk2}
is the Schwarzian derivative. The field redefinition $f=\tan(\phi/2)$ makes the action manifestly invariant under SL$(2,\mathbb{R})$ transformations
\eq{
f(\tau)\to\frac{a f(\tau)+b}{c f(\tau)+d}\,, \qquad ad-bc=1\,.
}{eq:syk3}
Hence, the Schwarzian action \eqref{eq:syk1} is not invariant under all reparametrizations of $\tau$, but realizes non-linearly the SL$(2,\mathbb{R})$ transformations \eqref{eq:syk3} due to the invariance of the Schwarzian derivative \eqref{eq:syk2} under fractional linear transformations. Thus, the infinite-dimensional symmetry group of diffeomorphisms of the circle (the Virasoro group) is broken to the finite-dimensional subgroup SL$(2,\mathbb{R})$. 

The gravity side of the holographic SYK story is described by the Jackiw--Teitelboim (JT) model \cite{Jackiw:1984, Teitelboim:1984}.  It is of interest to consider various extensions of SYK, since this enlarges the theory-space of possible holographic relationships and thus may allow to address relevant conceptual questions, for instance how general holography is and what are necessary ingredients for it to work. 

While these are intriguing questions, our goals in the present work are more modest, namely to supply candidates on the gravity side that generalize the symmetry breaking mechanism in SYK. 
In a sense, our approach is complementary to recent work by Gross and Rosenhaus \cite{Gross:2017vhb}, who considered free Majorana fermions in the large $N$ limit and conjectured that the bulk dual is some topological cousin of AdS$_2$ Vasiliev theory \cite{Bengtsson:2005zj, Fradkin:1989uh, Fradkin:1989kx, Li:1990nt, Li:1990wx, Vasiliev:1995sv, Rey:2011, Alkalaev:2013fsa, Grumiller:2013swa, Alkalaev:2014qpa, Mezei:2017kmw}: they worked on the field theory side [deforming the free theory by a bi-local bi-linear interaction preserving SL$(2,\mathbb{R})$], while our current paper deals exclusively with the bulk side (not necessarily related to the Gross--Rosenhaus model).

More specifically, our focus is to extend the symmetry breaking mechanism summarized above to other infinite-dimensional symmetry groups that contain a Virasoro subgroup. 

We are interested in two types of generalizations, one that has an interpretation in terms of dilaton gravity coupled to Yang--Mills and the other where Virasoro gets extended to $W$-symmetries, which arise in higher spin generalizations \cite{Alkalaev:2013fsa, Grumiller:2013swa, Alkalaev:2014qpa} of JT. Thus, our paper is aimed to provide the first few steps towards a higher spin (and Yang--Mills) generalization of SYK.\footnote{%
For additional work related to the SYK model and some of its generalizations see e.g.~\cite{Polchinski:2016xgd, You:2016ldz, Jevicki:2016bwu, Jensen:2016pah, Maldacena:2016upp, Engelsoy:2016xyb, Bagrets:2016cdf, Garcia-Alvarez:2016wem, Jevicki:2016ito, Gu:2016oyy, Gross:2016kjj, Berkooz:2016cvq, Garcia-Garcia:2016mno, Banerjee:2016ncu, Fu:2016vas, Witten:2016iux, Cotler:2016fpe, Klebanov:2016xxf, Davison:2016ngz, Peng:2016mxj, Krishnan:2016bvg, Turiaci:2017zwd, Ferrari:2017ryl, Bi:2017yvx, Li:2017hdt, Gurau:2017xhf, Mandal:2017thl, Gross:2017hcz, Mertens:2017mtv, Krishnan:2017txw}.
This list of references is necessarily incomplete, and we apologize for omissions.
}

The holographic dual description of a finite temperature quantum field theory is generated by placing a Euclidean black hole in the bulk. Let us suppose the set of black hole solutions preserves a certain (in lower dimensions typically infinite-dimensional) asymptotic symmetry group $\cG_{\infty}$. Demanding smoothness of the solutions yields a subset thereof that is invariant only under a subgroup $\cG\subset\cG_{\infty}$. In the SYK context this reproduces the symmetry breaking $\cG_{\infty}\to\cG$. The dynamics of the breaking is governed by a field belonging to the quotient space $\cG_{\infty}/\cG$. For instance, in the case of the Schwarzian action, the group $\cG_{\infty}$ is ${\rm Diff}(S^1)$, while $\cG={\rm SL}(2,\mathbb{R})$. The field $\phi$ is a diffeomorphism associated to the orbit ${\rm Diff}(S^1)/{\rm SL}(2,\mathbb{R})$ \cite{Stanford:2017thb}. The details of this construction (in first order formulation) were worked out in \cite{Grumiller:2017qao}.

In this paper we consider generalized models of dilaton gravity based on a gauge group $\cG$ \cite{Isler:1989hq, Chamseddine:1989yz}.  These theories do not propagate bulk degrees of freedom and are thus inherently ``holographic'' in the same way as three-dimensional Chern--Simons theories \cite{Witten:1989hf, Elitzur:1989nr}: physical excitations can be interpreted as edge states ``living at the boundary'' (see for instance \cite{Balachandran:1994up}). The suitable extension of the Schwarzian dynamics is governed by one-dimensional actions located at the boundary of the space-time. In order to extend the results of \cite{Grumiller:2017qao}, we study cases where $\cG$ contains an ${\rm SL}(2,\mathbb{R})$ subgroup. More precisely, we are interested in two cases:  Direct product groups, ${\rm SL}(2,\mathbb{R})\times \cK$, where $\cK$ is a compact group representing (Yang--Mills) matter fields and higher rank groups ${\rm SL}(N,\mathbb{R})$, where the spin two excitation is enhanced by the coupling of $N-1$ higher spin fields, analogous to the situation in three spacetime dimensions \cite{Henneaux:2012ny, Campoleoni:2012hp}. 

\enlargethispage{0.5truecm}

In order to construct these reduced models we follow two approaches. The first one evaluates the on-shell action. To do so, we construct a well-defined variational principle by adding boundary terms respecting the required symmetries. While this approach leads to unique generalizations of the Schwarzian action for direct product groups we are presented with difficulties in the case of higher spin extensions. Therefore, we find it more convenient to turn to an alternative, second approach to study higher rank groups $\textrm{SL}(N)$. This one analyzes the symplectic structure of dilaton gravity using the Lee--Wald--Zoupas formalism \cite{Lee:1990nz,Wald:1999wa}. Having a symplectic form allows us to define a geometric action that coincides with previous constructions \cite{Gegenberg:1997de} for JT. We study the inclusion of Hamiltonians that preserve the initial symmetry of the group $\cG$. We find Schwarzian-type of actions for higher spin fields by using the relation between holonomy conditions, that guarantee the smoothness of the associated geometric solution, and certain ordinary differential equations, that are well-known in the context of $W$-algebras.
Finally, we derive an entropy formula (including log-corrections from 1-loop effects) applicable to higher spin black holes in two dimensions and check its validity by recovering Wald's formula for the classical black hole entropy in the spin-2 case.

This paper is organized as follows.
In section \ref{se:2} we give a brief summary of dilaton gravity as non-abelian BF-theory, to fix our conventions, to clarify the type of observables we consider, and to formulate our action principle.
In section \ref{se:3} we reconsider the Jackiw--Teitelboim model to recover the Schwarzian action \eqref{eq:syk1}.
In section \ref{se:4} we consider Yang--Mills extensions of SYK.
In section \ref{se:5} we address higher spin extensions of SYK.
In section \ref{se:6} we study the symplectic structure of (generalized) dilaton gravity and present the generalization of the Schwarzian action for higher spin dilaton gravity. We also discuss the one-loop contribution of higher-spin fields to the partition function. 
In section \ref{se:entropy} we derive the entropy for BF-theories and compare with Wald's formula for the black hole entropy.
In section \ref{se:7} we summarize our results in the form of a holographic dictionary.
Appendix \ref{app:C} displays the quadratic and cubic Casimirs for spin-3 dilaton gravity.

\section{Dilaton gravity as generalized BF-theory}\label{se:2}

Generic dilaton gravity models \cite{Russo:1992yg, Odintsov:1991qu, Banks:1990mk, Mann:1990gh} can be reformulated in special cases as gauge theories \cite{Isler:1989hq, Chamseddine:1989yz, Cangemi:1992bj, Achucarro:1992mb} and more generally as non-linear gauge theories \cite{Ikeda:1993fh, Ikeda:1993aj} known as Poisson-$\sigma$ models \cite{Schaller:1994es}. These models are topological (see \cite{Birmingham:1991ty} for a review on topological quantum field theories) and rigid \cite{Izawa:1999ib} in the sense of Barnich and Henneaux \cite{Barnich:1993vg}, i.e., their most general consistent deformation is another Poisson-$\sigma$ model. See \cite{Grumiller:2002nm, Grumiller:2006rc} for review articles on two-dimensional dilaton gravity.

Like for Chern--Simons theories, not every Poisson-$\sigma$ model has a gravity-like interpretation, so for our purposes it is insufficient to merely write down some Poisson-$\sigma$ model bulk action. We need additionally a map from the gauge theoretic variables to gravitational entities (this point was emphasized in \cite{Strobl:2003kb}), which works most easily through the Cartan formulation of the latter. In other words, within our set of gauge connections we need to identify the combinations corresponding to zweibein and Lorentz connection. A sufficient requirement for a gravity-like interpretation analogous to the JT model is the existence of an SL$(2,\mathbb{R})$ sector in the gauge algebra, as we shall review below. Since this sector is linear, i.e., allows a simpler interpretation of the Poisson-$\sigma$ model as non-abelian BF-theory, we are going to consider exclusively extensions of JT that preserve linearity in the present work. 

The second ingredient for a satisfactory gravity interpretation, particularly in a holographic context, is the imposition of suitable boundary conditions on all fields, usually inspired by a certain design of the corresponding metric near the asymptotic boundary and by consistency requirements, such as the existence of a well-defined variational principle, see \cite{Grumiller:2017qao} for a menagerie of boundary conditions for the JT model. We stress already now one particular aspect of the boundary conditions that we are going to choose: we allow arbitrary leading order fluctuations of the dilaton near the asymptotic AdS$_2$ boundary, which generalizes many previous approaches towards AdS$_2$ holography. A fluctuating dilaton implies that a linear dilaton vacuum can be consistent with asymptotic AdS$_2$ isometries.\footnote{%
Holographic aspects of two-dimensional dilaton gravity were discussed in numerous papers, see \cite{Strominger:1998yg, Maldacena:1998uz, Cadoni:1999ja, Brigante:2002rv, Astorino:2002bj, Verlinde:2004gt, Grumiller:2007ju, Gupta:2008ki, Alishahiha:2008tv, Sen:2008vm, Hartman:2008dq, Castro:2008ms, Balasubramanian:2009bg, Castro:2009jf, Almheiri:2014cka, Cvetic:2016eiv} for a selected list of references. However, all these constructions start either with constant dilaton vacua, which are of less interest for SYK, or assume a fixed leading order component for the dilaton near the AdS$_2$ boundary, which restricts the space of solutions. The construction in the present work is instead based on linear dilaton vacua where the dilaton is allowed to fluctuate to leading order, as in \cite{Grumiller:2013swa, Grumiller:2015vaa, Grumiller:2017qao}.
} 

In the remainder of this section we review the non-abelian BF-formulation of JT (and generalizations thereof) in section \ref{se:2.1}, identify the relevant observables in section \ref{se:2.2} and formulate a well-defined action principle in section \ref{se:2.3}.

\subsection{Bulk action and equations of motion}\label{se:2.1}

\begin{figure}
\begin{center}
\begin{tikzpicture}
\draw[thick, ->] (-4,2.7) arc [start angle=320, end angle=30, x radius=0.2cm, y radius=0.4cm];
\draw (-5,2.9) node {}  node { $\tau$ };
\draw [thick,->] (-1,5.5)--(1,5.5) ;
\draw (0,6) node {}  node { $\rho$};
\draw (-4,2) node {}  node { \footnotesize $\rho=0$};
   
\draw[] (0,2.8) node {\includegraphics[width=7.8cm,height=4.4cm]{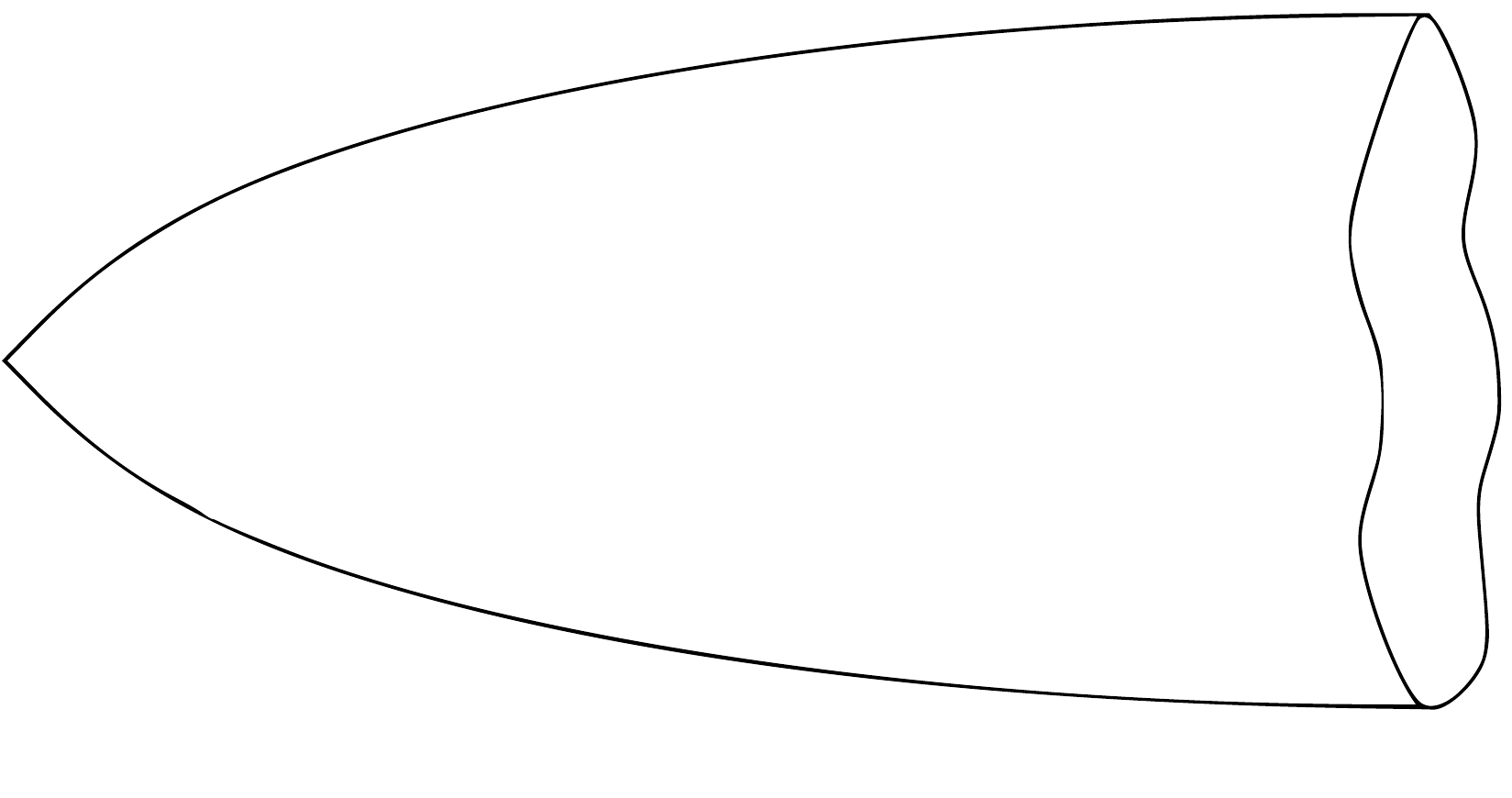}};      
\draw (5.3,3.4) node {}  node {\footnotesize $\mathcal{A}=\mathcal{A}_{\infty}(\tau)+\cdots$ }; 
\draw (5.3,2.4) node {}  node {\footnotesize $\mathcal{\cX}=\mathcal{\cX}_{\infty}(\tau)+\cdots$ }; 
\draw (-4.2,4) node {}  node {\footnotesize  ${\rm H}[\mathcal{A}]\neq\pm\unity$};
\end{tikzpicture}   

  \begin{tikzpicture}
  \centering
\draw[thick, ->] (-4,2.7) arc [start angle=320, end angle=30, x radius=0.2cm, y radius=0.4cm];
\draw (-5,2.9) node {}  node { $\tau$ };
\draw (-4,2) node {}  node { \footnotesize $\rho=0$}; 
\draw[] (0,2.8) node {\includegraphics[width=7.8cm,height=4.2cm]{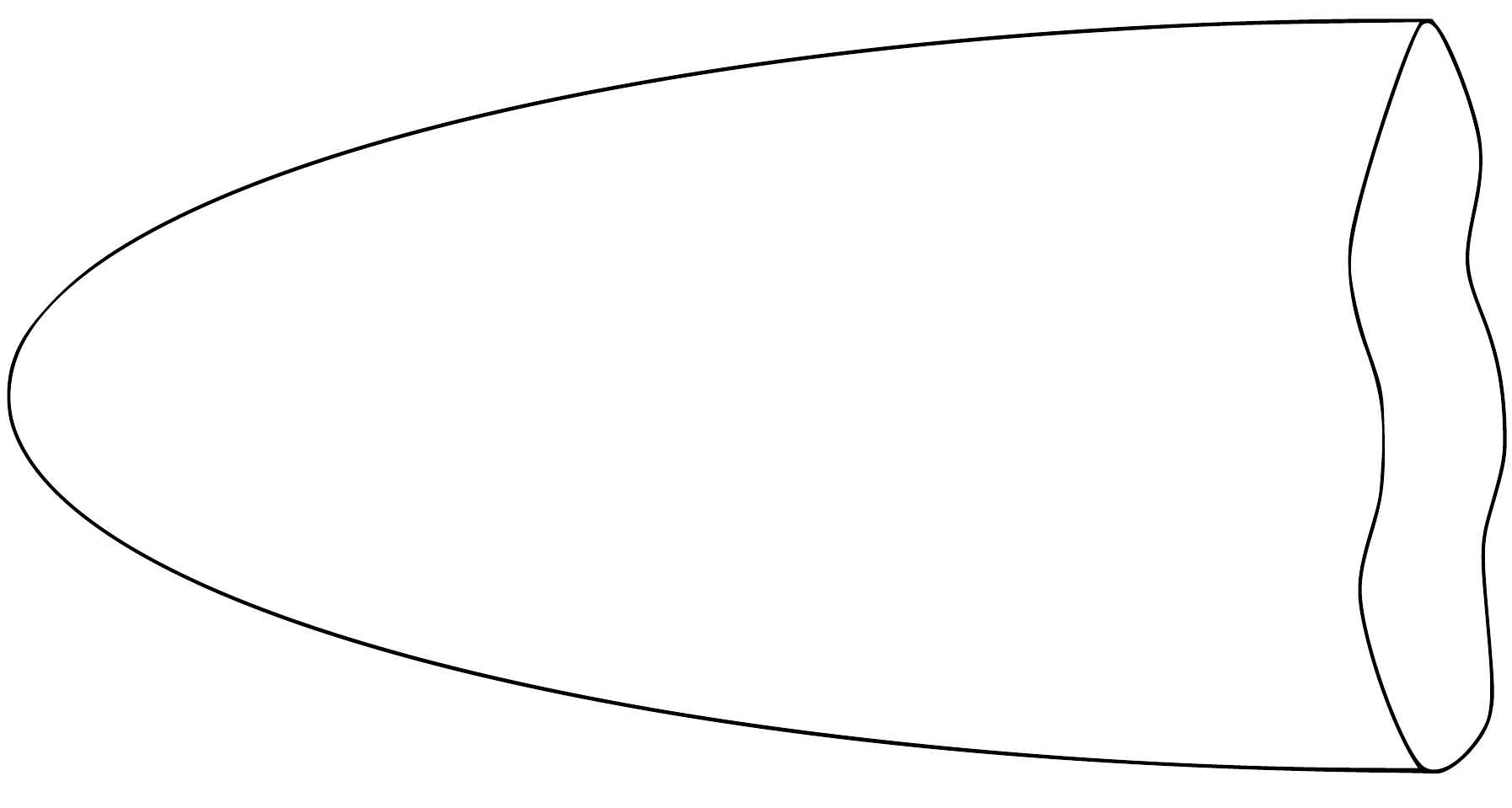}};      
\draw (5.3,3.4) node {}  node {\footnotesize $\mathcal{A}=\cA_{G}(\tau)+\cdots$}; 
\draw (5.3,2.4) node {}  node {\footnotesize  $\mathcal{X}=\cX_{G}(\tau)+\cdots$ }; 
\draw (-4.2,4) node {}  node {\footnotesize  ${\rm H}[\mathcal{A}]=\pm\unity$};
\end{tikzpicture}  
\end{center} 
\caption{ Finite temperature and asymptotic symmetry in the gauge theory formulation. Euclidean black holes are represented by fields $(\cA,\cX)$  in a cigar-type geometry. The ``Euclidean horizon'' is located at $\rho=0$. Demanding the black hole to be at Hawking temperature (absence of holonomies, ${\rm H}[\mathcal{A}]=\pm\unity$) affects the asymptotic symmetries. The asymptotic fields $(\cA_{\infty}(\tau),\cX_{\infty}(\tau))$ 
become $(\cA_G(\tau),\cX_G(\tau))$ consistently with smoothness of the 
solutions. 
} 
\label{fig:1}
\end{figure}
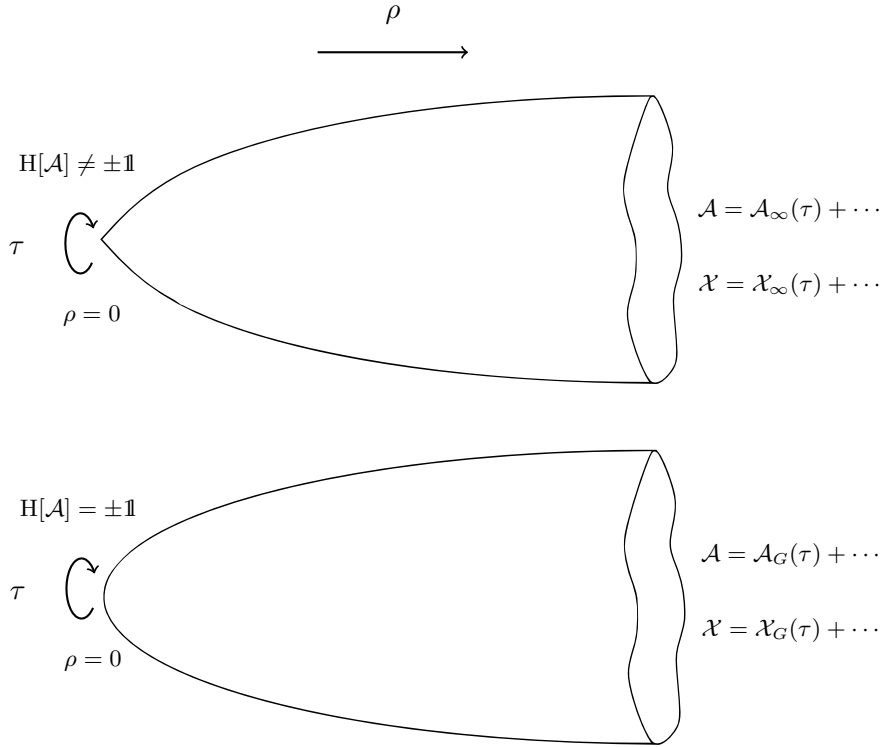

The gauge theoretic bulk action for (generalizations of) JT
\eq{
I_0[\cX,\cA]=\frac{k}{2\pi}\,\int\, \langle \cX, \cF \rangle
}{ac1}
contains the coadjoint ``dilaton'' $\cX=\cX^A J_A$, the non-abelian gauge field $\cA=\cA_{\mu}^A J_A \extd x^{\mu}$ and the associated curvature two-form $\cF=\extd\cA + \cA\wedge \cA$. 
Both fields are valued in the Lie algebra $\mathfrak{g}$ with generators $J_A$ satisfying $[J_A,J_B]=f_{AB}{}^CJ_{C}$ where $f_{AB}{}^C$ are the structure constants of $\mathfrak{g}$. We raise and lower algebra indices with the invariant metric $h_{AB}=\langle J_A , J_B \rangle$. The coupling constant $k$ is related to the two-dimensional Newton constant $G$ by $k=1/(2G)$. 
Throughout this work we consider fields that live in a space with the topology of a disk endowed with coordinates $(\tau, \rho)$ whose ranges are $0<\rho<\infty$ and $\tau\sim\tau+\beta$. For details see figure \ref{fig:1}.

The BF-theory \eqref{ac1} is gauge invariant. Given a Lie algebra parameter $\epsilon$, the fields transform as
\eq{
\delta_{\epsilon} \cA= \extd \epsilon+[\cA,\epsilon] \qquad\qquad  \delta_{\epsilon} \cX=[\cX,\epsilon]
}{ac2}
and the infinitesimal variation of the bulk action \eqref{ac1} becomes a boundary term.  The field equations that are obtained by varying \label{actm} with respect to $\cX$ and $\cA$ are  
\eq{
 \cF=0 \qquad\qquad d\cX + [\cA, \cX]=0\,.
}{ac3}
The first equation tells us that the on-shell connection is pure gauge, $\cA=-(\extd G) G^{-1}$, with $G \in \cG$ a not necessarily single-valued group element (that may account for non-trivial holonomies). The dynamics of the dilaton corresponds exactly to a gauge transformation that preserves the form of $\cA$ or, in other words, $\cX_{\rm on-shell}$ is the stabilizer of $\cA$.

By our assumption, $\mathfrak{g}$ must contain an $\mathfrak{sl}(2,\mathbb{R})$ subalgebra, which then allows an identification of this $\mathfrak{sl}(2)$-part with Cartan variables (zweibein and dualized Lorentz connection), see \cite{Grumiller:2015vaa, Grumiller:2017qao} for details. This provides the first necessary ingredient for a gravity-interpretation of the non-abelian BF-theory \eqref{ac1}. The second ingredient, specific boundary conditions on the connection and the dilaton, are provided in section \ref{se:3} below for the JT model and in later sections for generalizations thereof.

\subsection{Observables}\label{se:2.2}

We can construct two types of observables for BF-theories \eqref{ac1}: Wilson loops around the $\tau$-cycle and Casimir functions. 

The former are expressed as
\eq{
{\rm H}[\cA]= \cP \exp\Big[- \oint \cA \Big]
}{eq:angelinajolie}
where $\cP$ denotes path ordering and the integral is over the $\tau$-cycle whose period is $\beta$. For pure gauge connections we have $\extd G+ \cA G=0$, so solving $G$ in terms of $\cA$ yields ${\rm H}=G(\beta)G(0)^{-1}$. One is forced to demand that ${\rm H}$ belongs to the center of the group in order to single out smooth Euclidean solutions, yielding $G(\beta)= \cZ G(0)$ where $\cZ$ commutes with all the elements of $\cG$. In the case of  $\cG={\rm SL}(N,\mathbb{R})$ one chooses $\cZ=(-1)^{N+1}\unity$ as element of the center.

Another important class of gauge invariant observables are Casimir functions. Any semi-simple Lie-algebra $\mathfrak{g}$ admits invariant tensors $g_{A_1 \cdots A_n}$, where $n$ ranges from two to $1+$rank of $\mathfrak{g}$ [for $\mathfrak{sl}(N)$ the range is from two to $N$]. The associated Casimirs are defined as
\eq{
\cC_{n}=-\frac{1}{n}g_{A_1\cdots A_n} \cX^{A_1}\cdots\cX^{A_n}\,.
}{casn}
Casimir functions play the role of conserved charges of the theory. Indeed, the dilaton equation of motion \eqref{ac3} establishes the conservation equations
\eq{
\d_{\tau} \cC_n=0\,. 
}{eq:syk4}
In the simplest case of $\mathfrak{sl}(2)$ the rank is 1, the Casimir is interpreted as black hole mass and the conservation equation \eqref{eq:syk4} implies energy conservation.

\subsection{Action principle}\label{se:2.3}

We provide now a well-defined action principle for the space of solutions relevant in the context of two-dimensional dilaton gravity and generalizations thereof. Note first that without boundary terms the action \eqref{ac1} does not have a well-defined variational principle: infinitesimal variation yields
\eq{
\delta I_0={\rm(bulk~equations~of~motions)}+ \frac{k}{2\pi}\,\int_{\rho=\infty} \!\!\!\!\!\!\extd\tau\; \langle \cX, \delta \cA_{\tau} \rangle\, .
}{act0}
The last term in \eqref{act0} spoils the variational principle. However, we can get rid of the last term by adding a suitable boundary term $I_B$ to the bulk action \eqref{ac1}.  Demanding 
\eq{
\delta I\big|_{\textrm{\tiny EOM}} = \delta I_0\big|_{\textrm{\tiny EOM}} + \delta I_B\big|_{\textrm{\tiny EOM}} \stackrel{!}{=} 0
}{eq:syk68}
we find the following consistency condition
\eq{
\delta I_B = -\frac{k}{2\pi}\, \int_{\rho=\infty}\!\!\!\!\!\! \extd\tau \; \langle \cX, \delta \cA_{\tau}\rangle\,.
}{act2}
In order to find a local expression for $I_B$, we need pull the variation $\delta$ out of the integral. Without further assumption this cannot be done. To resolve this issue an integrability condition, 
\eq{
\cA^\infty_\tau=f(\cX^\infty)\,,
}{act3}
is needed. Here $f$ is an arbitrary function of the dilaton $\cX$ and the superscripts $\infty$ denote evaluation of the corresponding quantity in the limit $\rho\to\infty$. By means of the integrability condition \eqref{act3} we can, in principle,  find a local expression for $I_B$. In the next sections we examine different examples and provide integrability conditions \eqref{act3} respecting the symmetries of the problem. 

In order to choose our boundary conditions we follow the ideas of \cite{Henneaux:2013dra}. This amounts to pick a connection satisfying certain asymptotic conditions associated with a group $\cG_{\infty}$. The dilaton field is chosen to be the gauge parameter that preserves the form of the gauge field. This choice naturally selects an integrability condition that allows to define a well-defined variational principle.

\section{Jackiw--Teitelboim model}\label{se:3}

The JT model is obtained as non-abelian BF-theory described in section \ref{se:2} by choosing as gauge group $\cG={\rm SL}(2,\mathbb{R})$. The invariant tensor is determined by the matrix trace $\langle L_m, L_n \rangle =\tr[L_nL_m]$, and the generators $L_m \in \mathfrak{sl}(2,\mathbb{R})$ with $m=\{-1,0,1\}$ satisfy the usual commutation relations $[L_m,L_n]=(m-n)L_{m+n}$. The fundamental representation for these generators is
\eq{
L_1=\begin{pmatrix}
   0 & 0  \\
   1 & 0 
  \end{pmatrix}\qquad 
  L_0= \frac{1}{2}\begin{pmatrix}
   1 & 0  \\
   0 & -1 
  \end{pmatrix} \qquad 
  L_{-1}= \begin{pmatrix}
   0 & -1  \\
   0 & 0
  \end{pmatrix} \,.
}{eq:syk5}

\subsection{Boundary and integrability conditions}\label{se:3.1}

To specify boundary condition for the dilaton and the gauge field we employ a convenient parametrization of the fields \cite{Grumiller:2017qao}.  
\eq{
\cA= b^{-1} (\extd + a) b\qquad\quad  \cX= b^{-1} x(\tau) b \qquad\quad a= a_{\tau}(\tau) \extd\tau \qquad\quad  b=\exp(\rho L_0)
}{des}
We are interested here in boundary conditions for the gauge connection that asymptotically preserve Virasoro symmetries. They are most conveniently represented in the so-called highest weight gauge for the field $a_\tau$
\eq{
a_{\tau}=L_{1}+\cL(\tau)\, L_{-1}\,.
}{FGV}
In order to choose boundary conditions for the dilaton $x$ we proceed as follows. First, let us study the gauge symmetries that preserve the form of the auxiliary connection \eqref{FGV}. Solving 
\eq{
\delta_{\Lambda} a_{\tau} = \d_{\tau} \Lambda +[a_{\tau}, \Lambda]= \cO(a_{\tau})
}{sym}
yields
\eq{
\Lambda[\ve; a_{\tau}]=\ve L_1-\ve' L_{0}+\big(\cL \ve+\tfrac{1}{2}\ve''\big)L_{-1} 
}{eV}
and implies transformation of the function $\cL$ by an infinitesimal Schwarzian derivative.
\eq{
\delta_{\ve} \cL=\ve \cL'+2 \ve' \cL+ \tfrac{1}{2} \ve'''
}{eq:syk6}
As we saw in section \ref{se:2.1}, the dilaton field $x$ is the stabilizer of $a$. Thus it satisfies $\delta_x a_{\tau}=0$. We assume that $x$ has the form of \eqref{eV} with $\ve$ replaced by some $y$.
\eq{
x= \Lambda[y; a_{\tau}]
}{choiceV}
The on-shell value of the dilaton satisfies the relation $\delta_y \cL=0$. This condition corresponds to the little group equation of a Virasoro coadjoint orbit for the representative $\cL$ \cite{Witten:1987ty}.

The latter choice for $x$ has some nice consequences. The first one is that $y$ transforms as a one-dimensional vector field.  From the dilaton transformation \eqref{ac2} we have $\delta_{\eps} x= [\Lambda[y],\Lambda[\varepsilon]]$.\footnote{Notice that this is true only asymptotically, i.e., to leading order if the $\rho$ dependence is reinstated. Due to the implicit dependence of $x$ on $a_\tau$ through \eqref{choiceV}, it is the modified bracket 
$\delta_\eps x=[\Lambda[y],\Lambda[\eps]]_M=[\Lambda[y],\Lambda[\eps]]+\delta^a_y \Lambda[\eps]-\delta^a_\eps \Lambda[y]$ that closes as an algebra $[\Lambda[y],\Lambda[\eps]]_M=\Lambda[\eps y'-\eps'y]$. Here, $\delta^a_\eps \Lambda$ denotes the variation in $\Lambda$ under the variation of $a_{\tau}$ induced by $\eps$. } The component along $L_1$ of this expression tells us
\eq{
\delta_{\ve} y= \ve y'-y \ve'
}{eq:syk7}
which is the announced vectorial transformation. The second consequence is that the choice \eqref{choiceV} gives us a suitable integrability condition. In fact, we can reexpress \eqref{choiceV} in a simpler manner
\eq{
x= y\,\big(a_{\tau}-u^{-1} \d_{\tau} u\big)\,,
}{xa}
with  $u= \exp(-{\frac{1}{2}y' L_{-1}}) \exp(\log(y)L_0)$. By inverting this relationship we can express the gauge field as
\eq{
a_\tau=f_{\tau} x + u^{-1} \d_{\tau} u,
}{act4}
where $f_{\tau}=1/y$. From a more general perspective \eqref{act4} can be used as an integrability condition \eqref{act3} that relates the asymptotic connection $a_{\tau}$ with $x$ in terms of two quantities: a one-form $f_{\tau}\,\extd\tau$  and a group element $u$. They are free boundary data.  

Condition \eqref{act4} effectively appeared already in \cite{Grumiller:2017qao}, where the following expressions relating $a_{\tau}$ and $x$ were proposed [Eqs.~(3.5)-(3.7) of that reference]
\eq{
 \cL^{+}=\frac{1}{y}\cX^{+}\qquad\qquad
 \cL^{0}=\frac{1}{y} \cX^{0}+ \frac{(\cX^+)'}{\cX^+}\qquad\qquad
 \cL^{-}=\frac{1}{y} \cX^{-}+ \frac{(\cX^0)'}{2\cX^+}\,.
}{act4.5}
By choosing $a_\tau = \cL^{i} L_i$, $x= \cX^{i} L_i$, and $u= e^{\frac{1}{2}\cX^0 L_-} e^{\log(\cX^{+})L_0}$ we reproduce \eqref{act4.5} in the form of \eqref{act4}. We have thus succeeded in recasting the results \eqref{act4.5} in a gauge-covariant form \eqref{act4}, which will facilitate generalizations to higher spins and/or the inclusion of Yang--Mills.

\subsection{Action principle and on-shell action}\label{se:3.2}

Inserting our boundary condition \eqref{act4} into the variation of the boundary term \eqref{act2} one obtains
\eq{
\delta I_B=-\frac{k}{2\pi}\int \extd\tau \left[  \delta(f_{\tau}\, C )+ C \delta f_{\tau} -  \tr\left( (\d_\tau x+[u^{-1}\d_{\tau}u,x]) u^{-1} \delta u  - \d_{\tau} (x \,u^{-1} \delta u ) \right) \right]\,.
}{act5}
 Note that $f_{\tau}$ is a one-form component in one dimension, hence it can be written as $f_{\tau}=\frac{1}{\bar{y}}\d_{\tau} f$, where $1/\bar{y}$ is the zero mode of the quantity $1/y$ introduced in \eqref{xa}. We assume, additionally, that $f(\tau)$ is a well-defined diffeomorphism respecting $f(\tau+\beta)=f(\tau)+\beta$. This ensures that the second term vanishes since on-shell the Casimir $C$ is constant. The third term is zero on-shell and we can discard the last term by imposing that the fields are periodic on the $\tau$ cycle. 

From \eqref{act5} together with the parametrization for $\cX$ and $\cA$ explained in section \ref{se:3.1} we can infer that the bulk-plus-boundary action
\eq{
I[\cX,\cA]=I_0[\cX,\cA]+\frac{k}{4\pi \bar{y} }\int \extd\tau \, \big(\d_{\tau} f\big)\, \tr(\cX^2)
}{act6}
has a well-defined action principle. (Note that the term $\extd f = \extd\tau \, \big(\d_{\tau} f\big)$ acts as a boundary volume form.) Moreover, since the field strength vanishes on-shell, the corresponding value of the on-shell action is
\eq{
I_{\rm on-shell}= -\frac{k\beta}{2\pi\bar{y}}\,C
}{eq:syk10}
Taking into account the different normalizations of the BF coupling constants, $k_{\rm here}=2k_{\rm there}$, the result for the on-shell action \eqref{eq:syk10} agrees precisely with \cite{Grumiller:2017qao}. 

Comparing the overall normalization in the on-shell action \eqref{eq:syk10} with the one in the Schwarzian action \eqref{eq:syk1} it is plausible to relate the number $N$ of Majorana fermions holographically to the inverse Newton constant, and hence with $k$, like in usual holographic dictionaries where large $N$ corresponds to small Newton constant. Comparing (5.21) in \cite{Grumiller:2017qao} with (4.173) in \cite{Maldacena:2016hyu}, the strength $J$ of the random coupling gets identified with the zero mode $1/\bar y$. Thus, we have the following two entries in the SYK/JT holographic dictionary:
\eq{
N \sim k\qquad \qquad J \sim 1/\bar y
}{eq:syk11}
The low temperature condition then translates into the requirement 
\eq{
\bar y\ll\beta\qquad\Leftrightarrow\qquad \ell_f\gg 1
}{eq:syk12}
where $\ell_f=\frac{\beta}{\bar{y}}$ is the cycle of $f_{\tau}$ that effectively decompactifies. In the next subsection we make the holographic relation to SYK more precise by recovering the Schwarzian action \eqref{eq:syk1} on the gravity side.

\subsection{Boundary action from Casimir}\label{se:3.3}

The relation \eqref{xa} permits to express the Casimir in terms of $f_{\tau}= \frac{1}{\bar{y}} f'$ and $\cL$.
\eq{
C=\frac{\bar{y}^2}{f'^2}\,\big(\cL- \tfrac{1}{2}\{f\,;\tau\}\big)
}{act8}
Thus, the Casimir is determined from the coadjoint action of the Virasoro group. 

Under the diffeomorphism $u=f(\tau)$ and renaming $f^{-1}(u) \equiv \tau(u)$, the corresponding on-shell action \eqref{eq:syk10} reduces to the Schwarzian action \eqref{eq:syk1}.
\eq{
I_{\rm on-shell}[\tau]= -\frac{k \bar{y} }{2\pi}\int\limits^{\beta}_0 \extd u \, \big[ \tau'(u)^2\cL+\tfrac{1}{2} \{\tau;u\} \big]
}{act99}
Imposing regularity on the connection implies  ${\rm H}[a]=-\unity$. In order to satisfy this condition we need to provide the general solution to the equation $(\d_\tau + a_{\tau})g=0$ with anti-periodic boundary conditions $g(0)=-g(\beta)$ [Note that the relation between $G$, defined in section \ref{se:2.2}, and $g$ is $G=b^{-1}g$].

Using the connection in highest weight gauge, \eqref{FGV}, we find
\eq{
g=\begin{pmatrix}
   -\psi_1' & -\psi_2'  \\
   \psi_1 & \psi_2 
  \end{pmatrix}  
}{act10}
where $\psi_1$ and $\psi_2$ are two independent solutions to Hill's equation 
\eq{
(\d^2_\tau+\cL)\psi=0\,. 
}{eq:hills}
This equation transforms covariantly under the Virasoro group.\footnote{%
In the context of Liouville theory, classifications of Virasoro coadjoint orbits through Hill's equation have been analyzed in \cite{Balog:1997zz}. See also \cite{Oblak:2016eij} for a recent review. 
} If we make a diffeomorphism such that we go to the frame with constant $\cL$ we find solutions of the form $\exp(i \sqrt{\cL} \tau)$. In order to satisfy the anti-periodic boundary condition we find 
\eq{
\cL=\pi^2 n^2/\beta^2\qquad n\in\mathbb{Z} 
}{eq:syk59}
which corresponds to an element of the Virasoro coadjoint orbit whose little group is an $n$-cover of ${\rm SL}(2,\mathbb{R})$. (In most applications we set $n=1$.) Thus,  the value of $\cL$ in \eqref{act99} is restricted to be \emph{any} element of the orbit ${\rm Diff}(S^1)/{\rm SL}(2,\mathbb{R})$.

\section{Yang--Mills extensions of SYK}\label{se:4}

Non-abelian BF-theories with gauge group SL$(2, \mathbb{R})\times \cK$, where $\cK$ is some Lie group, provide a way to generalize the Schwarzian action \eqref{act99}, while still allowing a gravity-interpretation, namely as dilaton gravity coupled to Yang--Mills. (In the special case where $\cK=U(1)$ we recover dilaton gravity coupled to a Maxwell field.) The goal of this section is to discuss this generalization.

The gauge field and the dilaton are given by
\eq{
\cA= A^iL_i+\alpha^a I_a\qquad \qquad \cX=X^iL_i+ \chi^a I_a
}{u1}
where $I_a$ are the generators of the Lie algebra $\mathfrak{k}$ associated with $\cK$, which commute with the sl$(2,\mathbb{R})$ generators $L_i$ and satisfy $[I_a,I_b]=f_{ab}{}^c I_c$. The new and non-vanishing components of the invariant tensor are
\eq{
 \langle I_a,I_b \rangle=-2 g_{ab}
}{u2}
where $g_{ab}$ is the Cartan--Killing metric of $\mathfrak{k}$. The BF-action \eqref{actm} reads
\eq{
I=\frac{k}{2\pi}\int \langle \cX , \cF \rangle=\frac{k}{2\pi}\int \big(\tr[XF] + \tr_{\mathfrak{k}}[\chi\,\Phi] \big)
}{u3}
where the first term corresponds to the JT model (the trace with no subscript refers to the sl$(2)$-part of the algebra) and the second term to additional Yang--Mills fields (the trace with subscript $\mathfrak{k}$ refers to the $\mathfrak{k}$-part of the algebra). In terms of the Cartan--Killing metric the latter trace is given by
\eq{
 \tr_{\mathfrak{k}}[\chi\,\Phi]=-\, 2 \, g_{ab} \, \chi^a\, \Phi^b = -\, 2 \, g_{ab} \, \chi^a\,\big(\extd\alpha^b + f_{cd}{}^b\alpha^c\wedge\alpha^d\big)
}{eq:syk8}
where $\Phi$ is the Yang--Mills field strength.

To clarify our nomenclature, we refer to the gauge fields $A^i$ as ``gravitational'', to the gauge fields $\alpha^a$ as ``Yang--Mills'', and when addressing them together as ``gauge fields''. The coadjoint 0-forms $X^i$ and $\chi^a$ are jointly referred to as ``dilaton multiplet'' and separately as ``dilaton'' and ``B-fields'', respectively.

The rest of this section is organized as follows. 
In section \ref{se:4.1} we specify our boundary (and integrability) conditions.
In section \ref{se:4.2} we consider the boundary action principle and generalize the Schwarzian action.
In section \ref{se:4.3} we are more explicit about the interpretation of the BF-theory above in terms of dilaton gravity (\'a la JT) coupled to Yang--Mills.

\subsection{Boundary conditions}\label{se:4.1}

We follow the same strategy as for JT (see section \ref{se:3}) in order to specify the asymptotic behavior of $\cA$ and $\cX$. We assume that a similar decomposition as the one considered in \eqref{des} applies here. The fields $\chi$ and $\alpha$ are $\rho$-independent at the boundary.

Let us consider the following expression for the gauge fields
\eq{
a_{\tau}=L_1+\big(\cL-\cP^2\big)L_{-1}\qquad \qquad \alpha_\tau=\cP^a I_a
}{u5}
where $\cP \equiv g_{ab}\cP^a\cP^b$. The left equation is essentially the same highest weight ansatz as \eqref{FGV}, but with an additional Sugawara-contribution to the stress-tensor from the Yang--Mills currents $\cP^a$. The transformations preserving the expressions \eqref{u5} are given by
\eq{
\Lambda_{\textrm{sl}(2)}[\ve]=\ve L_1 +\big[ \big(\cL-\cP^2\big) \ve+\tfrac{1}{2} \ve'' \big]L_{-1} -\ve' L_0\qquad \qquad
\Lambda_{\mathfrak{k}}[\ve, \mu^a]= \big(\tfrac{1}{2}\mu^a+\ve \cP^a \big)I_a
}{u6}
such that we reproduce the infinitesimal transformation laws associated to a Virasoro Kac--Moody algebra. 
\eq{
\delta \cL=   g_{ab}  \cP^a (\mu^b)' +\epsilon \cL'+ 2 \epsilon' \cL  +\frac{1}{2} \epsilon'''
\qquad\qquad
\delta \cP = (\varepsilon \cP)'+ \frac{1}{2}\left(\mu'+ [\cP,\mu]\right)
}{u7}
We construct again the dilaton multiplet as stabilizer.
\eq{
x=\Lambda_{\mathfrak{sl}(2)}[y; a_{\tau}]\qquad \qquad \chi= \Lambda_{\mathfrak{k}}[y,\nu^a; a_{\tau}]
}{u5.5}

The on-shell dynamics of the dilaton multiplet is given by $\delta_{y,\nu} \cL=0$ and $\delta_{y,\nu} \cP=0$. By inverting these relations analogous to section \ref{se:3} we propose the following integrability conditions.
\eq{
a_{\tau}=f_{\tau}x+u^{-1}\d_{\tau}u\qquad\qquad \alpha_{\tau}= f_\tau \chi + \lambda^{-1} \d_{\tau} \lambda
}{u3.3}
Again, the 1-form $f_\tau\extd \tau$ and group elements $u, \lambda$ are free boundary data, and the gauge fields and dilaton multiplet are expressed in terms of them. This completes the specification of our boundary conditions for the Yang--Mills extension of JT.

\subsection{Boundary action}\label{se:4.2}

The on-shell action has the same form as before \eqref{act6} where the Casimir now reads
\eq{
C=-\frac{1}{2}\langle \cX , \cX \rangle=-\frac{1}{2}\,\tr[x^2]+g_{ab} \chi^a \chi^b=-\frac{1}{2}\,\big(\tr[x^2]+ \tr_{\mathfrak{k}}[\chi^2]\big)\,.
}{u4}

Comparing \eqref{u5.5} with \eqref{u3.3} we can further identify $y^{-1}= f_{\tau}$ and $\nu=-2 f_{\tau}^{-1} \lambda^{-1} \lambda' $. As before, assuming $f_{\tau}=\frac{1}{\bar{y}}\d_{\tau} f$ yields
\eq{
C=\frac{\bar{y}^2}{f'^2} \Big(\cL- \frac{1}{2} \{f ;\,\tau\} -  2\tr_{\mathfrak{k}}[\cP \lambda^{-1} \lambda'] +  \tr_{\mathfrak{k}}[(\lambda^{-1} \lambda')^2]\Big)\,.
}{u8}
This expression shows that the Casimir is represented as the coadjoint action of an element $\cL$ under the action of the Kac--Moody--Virasoro group given by a diffeormorphism $f(\vp)$ and a group transformation with parameter $\lambda$. For the abelian case $\cK=U(1)$, this is the transformation law of a warped-Virasoro stress energy tensor found in \cite{Afshar:2015wjm}.

In terms of $\tau$, the on-shell action \eqref{eq:syk10} reduces to
\eq{
I_{\rm on-shell}[\tau,\lambda]= -\frac{k\bar{y}}{2\pi}\int_{0}^{\beta} \extd u \, \left[ (\tau')^2  \cL+\frac{1}{2} \{\tau; u\} -  \tr_{\mathfrak{k}}[2\cP \lambda^{-1} \lambda'-(\lambda^{-1} \lambda')^2] \right]\,.
}{act9}
This is our desired generalization of the Schwarzian action \eqref{eq:syk1}. It is expected to be relevant for the low-energy description of SYK-like models with global symmetries \cite{Gross:2016kjj, Klebanov:2016xxf,Witten:2016iux,Yoon:2017nig}. The allowed values of $\cL$ and $\cP$  are the ones that solve the holonomy condition, 
\eq{
{\rm H}[a]=-\unity\qquad \qquad  {\rm H}[\alpha]=\unity_{\cK}
}{act9.1}
where $\unity_{\cK}$ is the identity element of $\cK$ (or some other suitable element of the center of $\cK$). Recalling discussion at the end of section \ref{se:4.2}, the first condition is now solved by an element of the Virasoro coadjoint orbit 
\eq{
\cL - \cP^2= \frac{n^2 \pi^2}{\beta^2}\,.
}{act9.2}
For the second condition in \eqref{act9.1}, we need to provide the solution of $(\d_{\tau}+ \cP)g_{\cK}=0$ with periodic boundary conditions $g_{\cK}(0)=g_{\cK}(\beta)$. The latter equation transforms covariantly under finite gauge transformations, so we can use this freedom to solve the equation in the frame where $\cP$ is a constant Lie algebra element. In that case, solutions are of the form $g_{\cK}=\exp(-\cP \tau)$, where the boundary condition imposes 
\eq{
\exp(\beta \cP)=\unity\,.
}{act9.3}
This equation states that the eigenvalues of $\exp(\beta \cP)$ are multiples of $2\pi i$. In the  $\cK=U(1)$ case, this condition simply reduces to 
\eq{
\cP=2\pi im\, T \qquad m\in\mathbb{Z}\,.
}{eq:syk80}

\subsection{Interpretation as dilaton gravity-Yang--Mills theory}\label{se:4.3}

Before considering generalizations to higher spins we summarize briefly the second order interpretation of the theory we have considered (in the higher spin case we cannot provide such an interpretation).

In the second order formulation the action for JT minimally coupled to BF-Yang--Mills reads (see \cite{Cangemi:1993bb} for the abelian case)
\eq{
I=\int\extd^2x \sqrt{g}\, \big(X (R + 2\ell^{-2}) + \tr_{\mathfrak{k}}[\chi (^\ast\!F-\cE)]\big)
}{eq:syk9}
where $X$ is now a scalar field, the dilaton (it is contained as specific component in what we called ``dilaton'' in the first order formulation, see \cite{Grumiller:2017qao} for details), $^\ast\!F=\varepsilon^{\mu\nu}\Phi_{\mu\nu}$ is the dualized non-abelian field strength, $\cE$ is the on-shell value of the dualized field strength (i.e., the ``color-electric'' field), and $\chi$ are (Lie-algebra valued) Lagrange multipliers enforcing the on-shell relation $^\ast\!F=\cE$.

The first order version of the action \eqref{eq:syk9} is of BF-type, with the following commutation relations
\begin{subequations}
\label{eq:syk50}
\begin{align}
[P_a,\,P_b]&=\epsilon_{ab}\,\Big(\frac{1}{\ell^2}J+\cE^A\,I_A \Big) 
& [P_a,\,J]&=\epsilon_a{}^bP_b \\
[I_A,\,I_B]&=f_{AB}{}^C I_C 
&[P_a,\,I_A]&=[J,\,I_A]=0\,.
\end{align}
\end{subequations}
The Jacobi identities hold provided $\cE^A f_{AB}{}^C I_C=0$, which is obeyed for $\cE^A\propto I^A$ (assuming totally antisymmetric structure constants). For vanishing $\cE$ the first line reproduces the gravitational sl$(2,\mathbb{R})$ (with AdS$_2$ radius $\ell$ made explicit), and the second line the Yang--Mills algebra, with no mixing between them. This is the situation we have considered in sections \ref{se:4.1}, \ref{se:4.2}. 

If instead $\cE^A=\cB\,I^A$ then the commutator of two translations $P_a$ yields not only the cosmological constant term, but receives an additional contribution proportional to the quadratic Yang--Mills Casimir $I^A I_A$, which is a magnetic-like modification of the translation algebra. This case is on-shell equivalent to ordinary dilaton--Yang--Mills theory
\eq{
I=\int\extd^2x \sqrt{g}\, \big(X (R + 2\ell^{-2}) + \tr_{\mathfrak{k}}[F^{\mu\nu}F_{\mu\nu}]\big)
}{eq:syk43}
subject to the charge superselection $^\ast\!F^A=\cB\,I^A$, i.e., the constant of motion corresponding to the quadratic Casimir built from the Yang--Mills field strength is determined from some fixed $\cB$. In other words, the constant of motion associated with the color-charge is converted into a parameter in the action \eqref{eq:syk9}.\footnote{This is similar to the set-up considered in \cite{Grumiller:2014oha} where the charge associated with a $\textrm{U}(1)$ symmetry was interpreted as playing the role of a cosmological constant.}

Thus, while strictly speaking the theories \eqref{eq:syk9} and \eqref{eq:syk43} are different, as we just discussed their difference is marginal. Also the difference between the cases $\cE=0$ \eqref{u3} and $\cE^A=\cB\, I^A$ \eqref{eq:syk9} is marginal, since the term linear in $\cE^A$ in the algebra \eqref{eq:syk50} is a trivial central extension and can be generated by shifting the generator $J\to J+\ell^2\cE^A I_A$. Thus, we expect no essential changes for the Schwarzian type of action \eqref{act9} in the presence of non-zero electric field $\cE$.

\section{Towards higher spin SYK}\label{se:5}

We consider now generalizations of JT to higher rank gauge groups, such as SL$(N,\mathbb{R})$, which contain some SL$(2,\mathbb{R})$ subgroup that we interpret as the gravity-part. For sake of concreteness we focus on SL$(3,\mathbb{R})$ in this section.

Generators of the $\mathfrak{sl}(3,\mathbb{R})$ algebra are given by $\mathfrak{sl}(2,\mathbb{R})$ generators $L_i$ and additional (spin-3) generators $W_m$ with $-2\le m \le 2$. Their algebra reads
\begin{align}
[L_i,L_j]&=(i-j)L_{i+j} \label{sl2com}\\
[L_i,W_m]&=(2i-m)W_{i+m}\\
[W_n,W_m]&=-\frac{1}{3}(n-m)(2n^2-nm+2 m^2-8)L_{n+m}
\end{align}
where we used some convenient normalization of $W_n$ to fix the overall factor in the last commutator.
The line of argument in this section is similar to the previous sections. In section \ref{se:5.bcs} we state our boundary conditions that are direct $\textrm{SL}(3)$ analogues of those discussed in section \ref{se:3.1}. We find again that the equation of motion for the dilaton field resembles the stabilizer equation for the gauge connection $a_\tau$. In section \ref{se:5.2} we determine the configurations that are compatible with the chosen temperature. In section \ref{se:5.3} we find that the form of $x$ does not allow to unambiguously define an integrability condition for $a_\tau$ as it did in the previous sections. Thus, while we are able to write down a generalization of \eqref{act6} some parameters in the action remain unspecified. The task of determining higher spin analogues of the Schwarzian action is taken up again (and accomplished) in section \ref{se:6} starting from a different perspective. 
\subsection{Spin-3 boundary conditions}
\label{se:5.bcs}
By analogy to spin-3 gravity in three dimensions \cite{Henneaux:2012ny, Campoleoni:2012hp} and to previous constructions in two-dimensional dilaton gravity \cite{Alkalaev:2013fsa, Grumiller:2013swa} we expect a $W_3$ symmetry to emerge for a suitable choice of the gauge field and dilaton multiplet. 

As in the spin-2 case we proceed in highest-weight gauge
\eq{
a_\tau =  L_1+\cL(\tau)\,L_{-1}+\frac{1}{4}\cW(\tau)\, W_{-2}\,.
}{aW}
Transformations preserving the form of this connection are given by  $\Lambda[\ve, \chi]=\Lambda_L^{i} L_i+ \Lambda_W^{m} W_m$ where $\Lambda_L^1=\ve$, $\Lambda_W^{2}=\chi$ and
\begin{subequations}
\label{hs7}
\begin{align}
\Lambda_L^0&=-\ve'\\
\Lambda_L^{-1}&=\cL \ve+\frac{1}{2}\ve''-2\cW \chi\\
\Lambda_W^1&=-\chi'\\
\Lambda_W^0&=2 \chi \cL+\frac{1}{2}\chi''\\
\Lambda_W^{-1}&=-\frac{2}{3} \chi \cL'-\frac{5}{3} \chi' \cL-\frac{1}{6} \chi'''\\
\Lambda_W^{-2}&=\chi\cL^2+\frac{1}{4}\ve \cW +\frac{7}{12} \chi' \cL'+\frac{1}{6} \chi \cL''+\frac{2}{3} \chi''\cL+\frac{1}{24} \chi''''\,.
\end{align}
\end{subequations}
The fields $\cL$ and $\cW$ transform as
\begin{align}
\label{hs7.1}
\delta_{(\ve,\chi)} \cL&=  \ve \cL'+2\cL \ve'+\frac{1}{2}\ve'''-2\chi \cW' -3\chi' \cW,\\
\label{hs7.1.2}\delta_{(\ve,\chi)} \cW&=\ve \cW' +3\ve' \cW +\frac{32}{3} \cL^2 \chi'+\frac{5}{3} \cL \chi'''+ \frac{16}{3} (\cL^2)' \chi + \chi'' \cL' - (\Lambda_W^{-1})''\,.
\end{align}
We can conveniently write the transformation parameters \eqref{hs7}  as 
\eq{
\Lambda[\ve,\chi; a_{\tau}]=\ve a_{\tau} +2\chi \left(a_{\tau}^2- \tfrac{1}{3}\,\unity\, \tr[a_{\tau}^2]\right)+\omega
}{cw3}
with
\eq{
\omega=-\ve' L_0 + \frac{1}{2}\ve'' L_{-1}
-\chi' \, W_1 
+ \frac{1}{2} \chi'' \,W_0 + \Lambda_W^{-1}\,  W_{-1}
-\frac{1}{4}\left((\Lambda_W^{-1})' - \cL \chi'' \right)W_{-2}
}{eq:syk18}
The dilaton is chosen as
\eq{
x=\Lambda[y,z; a_{\tau}]\,.
}{xW}
From the dilaton's transformation law we learn
\begin{align}
\delta_{(\ve, \chi)} y&=\ve y' - y \ve' +\frac{2}{3}( z \chi''' - z''' \chi) +z''\chi'-z'\chi''-
\frac{32}{3} (\chi z' - z \chi' ) \cL\\
\delta_{(\ve, \chi)} z&=\ve z' - 2 z \ve'-y \chi'  +2 y' \chi\,.
\end{align}
Under diffeomorphisms generated by $\ve$, the quantity $y$ transforms as a one-dimensional vector, while $z$ transforms as a two-tensor on the circle.

In sl$(3)$, we can define a quadratic and a cubic Casimir as
\eq{
C_{2}=-\frac{1}{2}\, \tr(x^2)\qquad\qquad C_{3}=-\frac{1}{3}\, \tr(x^3)\,.
}{eq:syk19}
We show now that they are conserved quantities.  Using our definition for the dilaton \eqref{xW}, we find that
\eq{
\d_{\tau} C_2= 4 y \, \delta_{(y,z)} \cL -4 z \,\delta_{(y,z)} \cW\qquad\qquad 
\d_{\tau} C_3= -8 \cQ \, \delta_{(y,z)} \cL - 12 \cR \, \delta_{(y,z)} \cW
}{cwc}
where $\cQ=\frac{1}{8}{\tr}(L_{-1}x^2)$ and $\cR=\frac{1}{32}{\tr}(W_{-2}x^2)$. Casimir conservation, $\d_\tau C_{2,3}=0$, follows from the dilaton's equation of motion
\eq{
\delta_{(y,z)} \cL=\delta_{(y,z)} \cW=0\,.
}{eq:syk20}
Explicit results for the Casimirs $C_2$ and $C_3$ expressed in terms of $y,z,\cL,\cW$ can be found in appendix \ref{app:C}.

\subsection{Regular \texorpdfstring{$\mathfrak{sl}(3)$}{sl(3)} gauge fields} \label{se:5.2}

Analogue to the $\textrm{SL}(2)$ case, smoothness of the gauge field $a_{\tau} \extd\tau$ is equivalent to existence of a solution to the equation
\eq{ 
\Big(\d_{\tau}+L_1 + \cL L_{-1}+\frac{1}{4} \cW W_{-2}\Big) g=0\qquad \qquad g(0)=g(\beta)\,.
}{eq:syk24.2}
In the fundamental representation of ${\rm SL}(3,\mathbb{R})$, the group element $g$ is given by
\eq{
g=\begin{pmatrix}
   \cD \psi_1 & \cD \psi_2 & \cD \psi_3 \\
   -\psi_1' & -\psi_2' & -\psi_3' \\
   \psi_1 & \psi_2 & \psi_3
  \end{pmatrix}\,,  
}{eq:syk24.3}
where $\cD = \d^2_{\tau}+2\cL$ is Hill's differential operator and the functions $\psi_{1,2,3}$ are three independent solutions of
\eq{
\psi'''+4\cL \psi'+ 2(\cW+\cL')\psi=0
}{eq:syk24.4}
with boundary condition $\psi(0)=\psi(\beta)$, $\psi'(0)=\psi'(\beta)$ and $  \psi''(0)=\psi''(\beta)$. This equation transforms covariantly under $W_3$ transformations \cite{Ovsienko:1990vh}. Using this freedom, we focus on the case where $\cL$ and $\cW$ are constant. This equation has solutions of the form  $\exp(\lambda \tau )$ where $\lambda$ satisfies $\lambda^3+4\cL\lambda+2\cW=0$. At the same time, boundary conditions imply that $\beta\lambda=2\pi i  n$ with $n\in\mathbb{Z}$. Compatibility between these two conditions imply
\eq{\cL=\frac{n^2 \pi^2}{\beta^2}, \qquad \cW=0.}{eq:syk24.5}
This orbit is invariant under ${\rm SL}(3,\mathbb{R})$ transformations. In fact, one can check that for the values \eqref{eq:syk24.4}, the infinitesimal transformations of the fields \eqref{hs7.1} and \eqref{hs7.1.2} are preserved by 
\eq{\ve = \exp\Big(\frac{2\pi i}{\beta} q \tau \Big), \quad \chi= \exp\Big(\frac{2\pi i}{\beta} p \tau\Big) }{eq:syk24.6}
where $-1\leq q \leq 1$, $-2 \leq p \leq 2$  are two integers. This is an eight-dimensional vector space generating the $\mathfrak{sl}(3)$ algebra under Lie brackets.

\subsection{Action principle for spin-3 generalization of JT}
\label{se:5.3}
Let us find the boundary term $I_B$ consistent with the existence of an integrability condition of the form \eqref{act3}. Unlike the previous cases, \eqref{xW} does not permit to express $a_{\tau}$ in terms of $x$. Therefore, we propose an $\mathfrak{sl}(3)$ extension of \eqref{act4}
\eq{
a_{\tau}=f^{(2)} x + \tfrac{1}{2}\,f^{(3)} \left(x^2- \tfrac{1}{3}\,\unity\, \tr(x^2)\right) + u^{-1} \d_{\tau} u\,.
}{ax3}
Using condition \eqref{ax3}, the  variation of the boundary term reads
\begin{align}
\delta I_{B}&=-\frac{k}{2\pi} \int_{\rho=\infty} \!\!\!\!\!\!\extd\tau\, {\tr}(x\delta a_{\tau})\\
&=-\frac{k}{2\pi} \int_{\rho=\infty}\!\!\!\!\!\!\extd \tau\, \Big[\delta( f^{(2)} C_{2}+ f^{(3)} C_{3} ) +\delta f^{(2)} C_{2} + \delta f^{(3)} C_{3}\Big]\,.
\end{align}
 Using that $C_2$ and $C_3$ are constants of motion and provided $f^{(2)}$ and $f^{(3)}$ do not have varying zero modes, the integrated expression for the boundary term is
\eq{
I_{B}=-\frac{k}{2\pi}\int_{\rho=\infty} \!\!\!\!\!\!\extd\tau \left( f^{(2)} C_{2}+ f^{(3)} C_{3} \right)\,.
}{eq:syk21}

Let us discuss how the condition \eqref{ax3} can be reached in a way that respects our boundary conditions \eqref{aW} and \eqref{xW}. We can use these restrictions to fix the functional form of ten parameters: $f^{(2)}$, $f^{(3)}$ and eight components of $\Theta\equiv u^{-1}\d_{\tau} u$. Compatibility with the highest weight gauge \eqref{aW}, imposes six conditions
\eq{
{\rm Tr}[J( a_{\tau}-L_1)]=0
}{eq:syk22}
where $J\in\{L_{-1},L_0, W_1, W_{-1}, W_{0}, W_{-2}\}$. This is not enough to completely determine all parameters, as we still need to provide extra four conditions. Two of them are used to define $\cL$ and $\cW$ in terms of $f^{(2)}$, $f^{(3)}$, $\Theta^{-1}_L$ and $\Theta^{-2}_W$. For the remaining components we have to impose further restrictions compatible with the $\mathfrak{sl}(2)$ case. This suggests that $f^{(2)}=y^{-1}+\cO(z)$, while $f^{(3)}=\cO(z)$. 
Apart from this leading order behavior, we cannot fully characterize parameters $f^{(2)}$ and $f^{(3)}$ in terms of $y$ and $z$ and therefore we are not able to obtain a boundary action along the lines of the previous sections. 

Thus, in order to obtain the spin-3 version of the Schwarzian action, we follow a different strategy that is described in the next section.

\section{Symplectic structure of generalized Jackiw--Teitelboim models}\label{se:6}

In this section we discuss an alternative way to obtain a boundary action associated to BF theories in two dimensions, namely by providing a suitable Hamiltonian action. This is interesting in its own right, but particularly useful in the higher spin case where the direct determination of the boundary action starting from the on-shell action was not successful (see section \ref{se:5} above). 

In section \ref{se:6.1} we recover the action on a group manifold as boundary action, which provides the basis for establishing the connection between asymptotically $\textrm{AdS}_2$ spacetimes and solutions to certain differential equations that generalize \eqref{eq:hills} and \eqref{eq:syk24.4} in \ref{se:6.2}. In section \ref{se:6.3} we present higher spin Schwarzian actions in the zero temperature case and propose a generalization to the finite temperatures in section \ref{se:6.finite}. Section \ref{se:one-loop} extends the arguments of \cite{Stanford:2017thb} to determine the one-loop contribution of higher spin fields to the partition function. 

\subsection{Boundary action on group manifolds}\label{se:6.1}

The kinetic term of this Hamiltonian action is obtained from the symplectic form associated to \eqref{ac1}. From \eqref{act0}, we know that  a general variation of the Lagrangian gives
\eq{
\delta L={\rm( equations~of~motions)}+ \frac{k}{2\pi}\,\int \langle \cX, \delta \cA \rangle\,.
}{eq:syk25}
Following \cite{Wald:1999wa}, the boundary term defines a symplectic structure at a given slice $\Sigma=\{\tau= {\rm constant} \}$
\eq{
\Omega_{\Sigma}=\frac{k}{2\pi}\, \int_{\Sigma} \delta  \langle x, \delta a \rangle.
}{ome}
 We have made use of the factorization \eqref{des} in order to express everything in terms of $\rho$-independent quantities.

Now we restrict ourselves to the on-shell phase space of the theory. This corresponds to the solutions of \eqref{ac3} given by
\eq{
a=-\extd g g^{-1}\qquad \qquad x=g x_{0} g^{-1}\qquad\qquad \extd x_0=0\,.
}{solution}
Plugging \eqref{solution} into \eqref{ome} we find
\eq{
\Omega_{\Sigma}=-\frac{k}{2\pi}\,  \delta  \langle x , \delta g\,g^{-1}  \rangle|_{\partial \Sigma}\,,
}{ome1}
i.e., the symplectic structure is given by a pure boundary term. We can interpret this boundary symplectic form in the following way. Consider the manifold of field configurations $(x=g x_0 g^{-1},g)$, which can be identified with the Lie group $\mathcal{G}$, and let $\Gamma=(x(s),g(s))$ denote a path on this space with dimensionless curve parameter $s$. The variation $\delta$ is then regarded as providing a differential $\delta g\equiv\frac{\dd}{\dd s}g(s) \dd s$ on this space, and \eqref{ome1} is a symplectic form on the group manifold. We can use this to define a geometric action (for more details regarding geometric actions and applications to lower-dimensional gravity, consult, e.g., \cite{Barnich:2017jgw}).\footnote{A similar action has been constructed in \cite{Alekseev:1988ce, Alekseev:1988vx} where $x$ is a fixed co-vector. In the present approach, $x$ is considered a field that is allowed to vary.} Consider a surface $\mathcal{N}$ in the group manifold, bounded by the curve $\Gamma=\partial\mathcal{N}$. Then a geometric action is given by
\eq{
I_{\rm geom}[x, g]=\int_{\mathcal{N}} \Omega_{\Sigma} =-\frac{k}{2\pi}\,   \int_{\Gamma} \extd s \; \left \langle x , \frac{\extd}{\extd s} g \, g^{-1}  \right \rangle\,.
}{eq:syk42}
Since the original symplectic form \eqref{ome1} was defined on the boundary of the spacetime $\mathcal{M}$, we can regard \eqref{eq:syk42} as the kinetic term for a boundary action of a generic BF model. 
 In order to give dynamics to this model we need to include a Hamiltonian preserving $\cG$-invariance. The Casimir functions \eqref{casn} naturally preserve the symmetry along $\Gamma$. The most general choice is
\eq{
\cH=\frac{k}{2\pi}  \sum_{i=2}^N \mu^{(i)}(s) \, C_{i}
}{eq:syk26}
where $\mu_{i}$ are some arbitrary functions and $N$ some integer depending on the gauge group (for SL$(N,\,\mathbb{R})$ this number is $N$). Then, the natural dynamical system for (generalized) dilaton gravity models follows from the reduced action principle
\eq{
I^{\cH}_{\rm geom}[x, g]=-\frac{k}{2\pi}\,   \int_{\Gamma} \extd s \; \left( \left \langle x , \frac{\extd}{\extd s} g \, g^{-1} \right \rangle - \sum_{i=2}^N \mu^{(i)} \, C_{i}\right) \,.
}{geoH}
Note that the on-shell action for this system is given by the sum of the Casimir functions.
The equation of motion for $x$ is given by 
\eq{
\left(\frac{\extd}{\extd s} g \, g^{-1}\right)_{A} = - \sum_{i=2}^N \mu^{(i)}g_{A A_2 \cdots A_i } x^{A_2} \cdots x^{A_i}\,.
}{eq:syk26.1}
Plugging this back in the action \eqref{geoH} we find 
\eq{
I^{\cH}_{\rm geom}=-\frac{k}{2\pi} \sum_{i=2}^N (i-1) \int_{\Gamma} \extd s \, \mu^{(i)} C_i\,.
}{geoH2}
This acquires the same form  for the previously known cases with one \eqref{eq:syk10} and two Casimir functions \eqref{eq:syk21}. In what follows we consider the more tractable case  where $\mu^{(2)}$ is the only non-vanishing function. In that case, 
\eq{
\int \extd s \, \cH=-\frac{k}{4\pi}\, \int \extd s \, \mu^{(2)}(s)   \langle x ,  x  \rangle\,.
}{eq:syk27}
From the above expression, we see that $\mu^{(2)}$ plays the role of an einbein. Defining $1/\bar{y}$ as the zero mode of $\mu^{(2)}$, we can always choose a new coordinate $\tau$ such that  $ \mu^{(2)} \extd s = \frac{1}{\bar{y}}\extd\tau$ where $\mu^{(2)}= \frac{1}{\bar{y}} \tau'(s)$. 

It is convenient to express the action in the second order formulation. This is achieved by eliminating the momenta $x$ using equation \eqref{eq:syk26.1}
\eq{
x= \frac{\bar{y}}{\tau'} \d_{s} g \, g^{-1} \quad\implies\quad I^{\cH}_{\rm geom}[g]=\frac{k \bar{y}}{4\pi} \, \int \extd\tau \langle  \d_{\tau} g g^{-1}, \d_{\tau} g g^{-1}  \rangle\,.
}{qua}
Let us specialize to the case $\cG= {\rm SL}(N,\mathbb{R})$. Expression \eqref{qua} corresponds to the action of a particle on ${\rm SL}(N,\mathbb{R})$ group manifold (for details related to $N=2$ see \cite{Dzhordzhadze:1994np,Heinze:2015oha}). In the context of dilaton gravity, this was already found in \cite{Brigante:2002rv}.

The model \eqref{qua} is invariant under multiplication by constant group elements both from the left and the right
\eq{
g \mapsto a \,g\, b\qquad a,b\in {\rm SL}(N,\mathbb{R})\,.
}{lrsym}
It is straightforward to find the canonical charges that generate the symmetries \eqref{lrsym}. The symplectic potential 
\begin{equation}
    \theta(g,\delta g)=\frac{k \bar{y}}{2\pi} \langle g^{-1}\d_\tau g \,,g^{-1}\delta g\rangle
\end{equation}
yields the symplectic structure 
\begin{equation}
    \Omega=\frac{k\bar{y}}{2\pi}\langle g^{-1}\delta(\partial_\tau g),g^{-1}\delta g\rangle
\end{equation}
for the model \eqref{qua}. As expected, this is identical to the symplectic structure \eqref{ome1} after imposing the equation of motion for $x$. 

The vectors in field space $\xi$ that are tangent to the flows generated by the symmetries \eqref{qua} are given by
\begin{equation}
\label{vecA}
    \xi_A =-A g \qquad \qquad
    \xi_B =-g B \qquad A, B\in \mathfrak{sl}(N,\mathbb{R}),
\end{equation}
respectively, where $A$ and $B$ are the Lie algebra elements corresponding to $a$ and $b$ via the exponential map. The canonical charge $Q$ generating the flow tangent to a vector field $\xi$ is given by Hamilton's equation
\begin{equation}
    \delta Q= i_\xi \Omega 
\end{equation}
In the present case one finds 
\begin{equation}
    Q^L_{A}=\left \langle A,\frac{k\bar{y}}{2\pi}\d_\tau g g^{-1}\right \rangle,
\qquad
\label{rcharge}
    Q^R_{B}=\left \langle B,\frac{k\bar{y}}{2\pi} g^{-1}\d_\tau g\right \rangle
\end{equation}
as generator of left and right symmetry, respectively. The Poisson brackets between the charges, read off from the symplectic structure,
\begin{equation}
    \{Q^L_A,Q^L_{A'}\}=\frac{2\pi}{k \bar{y}}Q^L_{[A',A]}, \quad
    \{Q^R_{B},Q^R_{B'}\}=\frac{2\pi}{k \bar{y}}Q^R_{[B',B]}, \quad
    \{Q^L_{A},Q^R_{B}\}=0 \label{mPbs}
\end{equation}
show explicitly that the symmetry algebra consists of two commuting copies of $\mathfrak{sl}(N,\mathbb{R})$. 

\subsection{Reduction to the gravitational sector}\label{se:6.2}

The action \eqref{qua} was derived using the gauge flatness condition \eqref{solution} without assuming any particular form of $a$. However, in order to make contact with the previous sections the connection cannot be arbitrary but should fulfill two requirements:
\begin{itemize}
    \item the geometry associated to $a$ is asymptotically $\textrm{AdS}_2$ (with fluctuating dilaton);
     \item $a$ is compatible with the temperature, i.e., has no conical deficits. 
\end{itemize}
We will postpone the second item to section \ref{se:6.finite} which means that we restrict ourselves to the \emph{zero-temperature} case in the following. 

The interpretation of $a$ as describing an asymptotically $\textrm{AdS}_2$ geometry is guaranteed if the connection is taken to be of the (highest-weight) form
\begin{equation}
\label{eq:congen}
    a_\tau=L_{1}+Q\,,
\end{equation}
where $[L_{-1},Q]=0$. Connections \eqref{FGV} and \eqref{aW} considered in the previous sections belong to this class. More generally, choosing the principal embedding $\mathfrak{sl}(2)\hookrightarrow \mathfrak{sl}(N)$, the adjoint representation of $\mathfrak{sl}(N)$ is decomposed in irreducible representations $\{W_m^{s}\}$, $s=3,\dots,n$ of the `spin-2 gravity' subalgebra \eqref{sl2com} with commutators
\begin{equation}
    [L_m,W_n^{(s)}]=((s-1)m-n)W_{m+n}^{(s)}\,. 
\end{equation}
The above requirement thus restricts the connection \eqref{eq:congen} to be of the form
\begin{equation}
\label{eq:aexp}
    a_\tau=L_{1}+\mathcal{L} L_{-1}+\sum_{i=3}^N \mathcal{W}_i\,  W^{(i)}_{1-i}\,.
\end{equation}
Note that these generators provide an orthogonal basis for the Lie algebra with respect to the Cartan-Killing metric. 

The $\textrm{SL}(N)$ element $g$ therefore has to obey the equation
\begin{equation}
\label{eq:flata}
    \Big(\partial_\tau + L_{1}+\mathcal{L}\, L_{-1}+\sum_{i=3}^n \mathcal{W}_i\,  W^{(i)}_{1-i}\Big)\, g=0\,,
\end{equation}
where the fundamental representation for the $\mathfrak{sl}(N)$ elements is assumed. We demonstrate now that this equation implies a parametrization of $g$ in terms of the independent solutions to an $N$-th order differential equation, analogous to the cases discussed in section \ref{se:3.3} and \ref{se:5.2}. 

In the fundamental representation the operator acting on $g$ in equation \eqref{eq:flata} is of the form
\begin{equation}
\label{eq:matrixop}
    \begin{pmatrix}
    \partial_\tau& -\sqrt{k_1} \cL& \alpha_3 \cW_3&\alpha_4\cW_4&\cdots&\alpha_{N-1}\cW_{N-1}&\alpha_N\cW_N\\
    \sqrt{k_1}           &   \partial_\tau& -\sqrt{k_2}\cL& \alpha_3 \cW_3&\cdots&\alpha_{N-2}\cW_{N-2}&\alpha_{N-1}\cW_{N-1}\\
    0&\sqrt{k_2}&\partial_\tau&-\sqrt{k_3}\cL&\cdots &\alpha_{N-3}\cW_{N-3}&\alpha_{N-2}\cW_{N-2}\\
      \vdots & \vdots &\vdots &\vdots &\cdots &\vdots&\vdots \\
     0&0&0&\sqrt{k_{N-2}}&\cdots&\partial_\tau&-\sqrt{k_{N-1}}\cL\\
     0&0&0&0&\cdots&\sqrt{k_{N-1}}&\partial_\tau
     \end{pmatrix}\,,
\end{equation}
where $k_i=2\sum_j(K^{-1})_{ij}$ and $K_{ij}$ is the Cartan matrix and $\alpha_s$ is some normalization for the higher spin charges that, although straightforward to determine, will not be important in the following.

Writing the group element $g$ in terms of $n$-dimensional row vectors 
\begin{equation}
\label{eq:gpar}
    g=\begin{pmatrix}
    \Psi_{N}\\\Psi_{N-1}\\\vdots\\ \Psi_{2}\\\Psi_{1},
    \end{pmatrix}
\end{equation}
the structure of the operator \eqref{eq:matrixop} allows to express the vectors $\Psi_2,\dots,\Psi_N$ in terms of $(N-1)$ derivatives of $\Psi_{1}$. Denoting the components of $\Psi_1$ by $\psi_1,\psi_2,\dots,\psi_N$ the solution of \eqref{eq:flata} thus boils down to solving the $n$-th order differential equation
\begin{equation}
    \label{eq:genHills}
    \psi_i^{(N)}+u_2 \psi_i^{(N-2)}+u_{3}\psi_{i}^{(N-3)}+\dots+ u_{N-1}\psi_i^\prime + u_N \psi_i=0\,.
\end{equation}
Notice the absence of a term proportional to $\psi_i^{(N-1)}$. This is related to the fact that $g$ has determinant equal one, as it is an $\textrm{SL}(N)$ element. The differential equation \eqref{eq:genHills} generalizes Hill's equation \eqref{eq:hills} and has $N-1$ independent solutions that can be identified with the $\psi_i$ for $1\le i\le N-1$. The remaining function $\psi_N$ is then obtained from the determinant condition $\det(g)=1$. 

The coefficient functions $u_i$ are monomials of derivatives of $\mathcal{L}$ and $\mathcal{W}_i$. However, it is easy to show that the coefficient $u_2$ is always given by
\begin{equation}
    u_2=\frac{N(N^2-1)}{6} \,\mathcal{L}\,. \label{eq:U2}
\end{equation}

The differential equation \eqref{eq:genHills} transforms covariantly under $W_N$ transformations. While the transformation under arbitrary finite $W$ transformations is not known apart from some specific cases (see e.g., \cite{Gomis:1994rz}), under the subgroup of reparametrizations of $\tau$, i.e., the Virasoro group, $\psi_i$ transforms as
\begin{equation}
    \psi_i(\tau)=\left(\frac{\dd t}{\dd \tau}\right)^{-\frac{N-1}{2}}\psi_i(t)\,\qquad \tau\rightarrow \tau(t). 
    \label{eq:diffHills}
\end{equation}
For an infinitesimal transformation $\tau\mapsto \tau+\epsilon(\tau)$ one finds
\begin{equation}
\label{eq:psiinf}
    \delta_\epsilon \psi_i=\epsilon \psi_i'-\frac{N-1}{2}\epsilon' \psi_i\,.
\end{equation}
The coefficients $u_i$ have a complicated transformation behaviour under reparametrizations but it is  straightforward to show that $u_2$ transforms as an anomalous two-tensor
\begin{equation}
    u_2(t)=u_2(\tau)\left(\frac{\dd \tau}{\dd t}\right)^2+\frac{N(N^2-1)}{12}\,\{\tau;t\}\,,
\end{equation}
as suggested by the observation \eqref{eq:U2}. 

Further properties of the differential equation \eqref{eq:genHills} have been intensively studied in the context of W-algebras and their relation to KdV flows and Gelfand-Dikii Poisson structures. Here we do not go into the details of these interesting developments but refer to the ample literature, see e.g.~\cite{DiFrancesco:1990qr} and references therein.

 We saw above that the general model \eqref{qua} has a global $\textrm{SL}(N)\times \textrm{SL}(N)$ symmetry under multiplication from the left and from the right \eqref{lrsym}. However, it is clear from \eqref{solution} or \eqref{eq:flata} that the global left symmetry of the model is broken if $g$ in \eqref{qua} is required to be a solution of that equation. On the other hand, multiplication of $g$ on the right by an $\textrm{SL}(N)$ element is still a symmetry. The action on the $\psi_i$ of this symmetry is immediately clear from their representation as a row vector in \eqref{eq:gpar}, i.e., it is the natural action of $\textrm{SL}(N)$ on an element of $\mathbb{R}^N$. 
It is convenient to introduce the following ratios
 \begin{equation}
     s_i=\frac{\psi_i}{\psi_N}\qquad 1\le i\le N-1\,.
 \end{equation}
since the determinant condition $\det g=1$ allows then to express $\psi_N$ as a function of the $s_i$'s. Notice that the $s_i$'s can be viewed as homogeneous coordinates on the $(N-1)$ dimensional real projective space $\mathbb{RP}^{N-1}$. The differential equation \eqref{eq:genHills} is therefore associated with a curve $\gamma(\tau)=(s_1(\tau),..., s_{N-1}(\tau))\in \mathbb{RP}^{N-1}$. The action of $\textrm{SL}(N)$ on the $\psi_i$'s then induces the action $\textrm{PSL}(N)$ on the $s_i$. For instance in the case $N=2$ one finds the transformation 
\begin{equation}
\label{eq:sl2mob}
    s_1\mapsto \frac{a s_1+b}{c s_1 +d}
\end{equation}
in accordance with \eqref{eq:syk3}.

\subsection{Higher spin Schwarzian actions (zero temperature)}\label{se:6.3}

We are now ready to present the key point of the argument that allows us to construct analogues of the Schwarzian action in the higher spin cases: It is possible to construct $(N-1)$ \emph{projective invariants} $I^{(r)}(s_i;\tau)$ with $r=2,\dots,N$ from the solutions $s_i$ of the differential equation \eqref{eq:genHills}. They are invariant under the projective action of $\textrm{SL}(N)$ on $s_i$ and transform as $r$-tensors under reparametrizations of $\tau$. In particular, for $r=2$ one finds
\begin{equation}
\label{eq:proj2}
    I^{(2)}(s_i;\tau)=u_2(\tau)\,,
\end{equation}
with the same anomalous transformation law under diffeomorphisms. We do not present the general algorithm, which can be found, e.g., in \cite{Forsyth:1902,Govindarajan:1994wm} but outline the calculation for $N=2$. In this case we reproduce Hill's equation \eqref{eq:hills} with $u_2=\cL$ which reads 
\begin{equation}
    s_1''\psi_2+2s_1'\psi_2'=0\, .
\end{equation}
 Differentiating this equation and using Hill's equation again yields
\begin{equation}
    (s_1'''-2s_1'\mathcal{L})\psi_2+3s_1''\psi_2'=0\,.
\end{equation}
This set of two differential equation for $\psi_2$ and $\psi_2'$ can have a non-trivial solution only if its determinant vanishes. We therefore find
\begin{equation}
    \mathcal{L}=\frac{1}{2}\left(\frac{s_1'''}{s_1'}-\frac{3}{2}\left(\frac{s_1''}{s_1'}\right)^2\right)=\frac{1}{2}\{s_1;\tau\}\,.
\end{equation}
Since the Schwarzian derivative is invariant under fractional linear transformations such as \eqref{eq:sl2mob} and transforms anomalously under reparametrizations of $\tau$, we have succeeded in finding $I^{(2)}(s_i;\tau)$ in the case $N=2$. 

By equations \eqref{eq:U2} and \eqref{eq:proj2} it is always possible to write $\cL$ in terms of $N-1$ functions $s_i$
\begin{equation}
    \cL=\cL[s_1(\tau),\cdots,s_{N-1}(\tau)]\,.
\end{equation}
Inserting $a=-\partial_\tau g g^{-1}$ in the action \eqref{qua} and using the fact that our basis of $\textrm{sl}(N)$ is orthogonal we find that the action can be rewritten as
\begin{equation}
\label{eq:schwarzfamily}
    I[g]=\kappa \frac{k\bar{y}}{2\pi}\int\extd \tau\, \mathcal{L},
\end{equation}
where 
\eq{
\kappa=\tr(L_1L_{-1})\,.
}{eq:kappa}
By the arguments in the previous paragraphs this action is manifestly invariant under $\textrm{SL}(N)$ transformations. For $N>2$ it describes the appropriate higher-spin analogue of the Schwarzian action. Indeed, in the case $N=2$ we reproduce
\begin{equation}
    I[g]=-\frac{k\bar{y}}{4\pi} \int\dd \tau\left( \frac{s_1'''}{s_1'}-\frac{3}{2}\left(\frac{s_1''}{s_1'}\right)^2\,\right),
\end{equation}
while for $N=3$ we find
\begin{equation}
\label{eq:schwarziansl2}
    I[g]=-\frac{2 k\bar{y}}{\pi} \int\dd \tau\left( \frac{f'''}{f'}-\frac{4}{3}\left(\frac{f''}{f'}\right)^2+\frac{e'''}{e'}-\frac{4}{3}\left(\frac{e''}{e'}\right)^2-\frac{1}{3}\frac{f''e''}{f'e'}\right),
\end{equation}
where $e=s_1'/s_2'$ and $f=s_2$. This action and its relation with $W_3$ algebras is well-known, e.g., \cite{Marshakov:1989ca, Li:2015osa}.  The $\textrm{SL}(3)$ invariance of this action, guaranteed by the above arguments, can be checked in a straightforward if tedious manner using the transformations
\begin{equation}
\label{eq:schwarziansl3}
    s_1\mapsto \frac{a_{11} s_1+ a_{12} s_2 +a_{13}}{a_{31} s_1+a_{32} s_2+a_{33}} \qquad\qquad   s_2\mapsto \frac{a_{21} s_1+ a_{22} s_2 +a_{23}}{a_{31} s_1+a_{32} s_2+a_{33}}\,,
\end{equation}
where $a_{ij}$ denote components of an $\textrm{SL}(3)$ matrix.

\subsection{Higher spin Schwarzian actions (finite temperature)}

\label{se:6.finite}

We deal now with the second item of the list at the beginning of section \ref{se:6.2}, i.e., we demand the connection $a$ to be compatible with the temperature given by the inverse of the periodicity of Euclidean time $\beta$. We set $\beta=2\pi$ in this subsection. 

As discussed in section \ref{se:2.2}, the absence of conical singularities is guaranteed if the connection \eqref{eq:aexp} has trivial holonomy. This is equivalent to the condition
\begin{equation}
    g(2\pi)=(-1)^{N+1}\, g(0)
\label{eq:holo}    
\end{equation}
on the group element \eqref{eq:gpar}. The solutions $\psi_i$ we are looking for therefore have to obey (anti-) periodic boundary conditions for $N$ (even) odd: $\psi_i(2\pi)=(-1)^{N+1}\psi_i(0)$. This is consistent with the fact that by equation \eqref{eq:psiinf} a solution $\psi_i$ has (half-)integer spin for (even) odd $N$. 

The holonomy condition on the group elements $g$ \eqref{eq:holo} can be reformulated in the following neat geometric way. We mentioned above that a solution to \eqref{eq:genHills} can be viewed as a curve in $\mathbb{RP}^{N-1}$ with Euclidean time $\tau$ as parameter. Since we are working at finite temperature $\beta$ this corresponds to a map $\gamma: S^1\rightarrow \mathbb{RP}^{N-1}$.

A generic solution $\gamma$ is not closed but is shifted by an element $M\in \textrm{SL}(N)$, called \emph{monodromy}, after one period of Euclidean time
\begin{equation}
    \gamma(\tau+2\pi)=M\gamma(\tau)\qquad\qquad M\in \textrm{SL}(N)\,. 
\end{equation}
The (conjugacy class of the) monodromy is an invariant of the differential equation \eqref{eq:genHills} \cite{Khesin:1990sy}. In other words, acting on the differential equation \eqref{eq:genHills} with an arbitrary $W$ transformation leads to a different solution curve albeit with the same monodromy as the solution curve of the original equation. The \emph{homotopy class} of curves with a given monodromy, i.e., the (in-)ability to deform one into the other, provides another invariant for the differential equation \eqref{eq:genHills}. In fact, monodromy and homotopy class are the only invariants \cite{Khesin:1990sy}. 

The holonomy condition \eqref{eq:holo} is thus translated to the following statement: To each solution curve $\gamma\in \mathbb{RP}^{N-1}$ one can associate its unique lift $\tilde{\gamma}=(\psi_1,\dots,\psi_N)\in \mathbb{R}^{N}$, as guaranteed by the determinant condition. The monodromy of the lift $\tilde{\gamma}$ has to obey 
\eq{
\widetilde{M}=(-1)^{N+1}\unity\,.
}{eq:Mg0}
This means that the monodromy of $\gamma$ is 
\eq{
M=\unity\,.
}{eq:Mg1}
The number of homotopy classes for curves $\gamma \in \mathbb{RP}^{N-1}$ of monodromy $\unity$ with the above properties have been determined in \cite{Shapiro:1992to} (see also \cite{Bajnok:2000nb,Ovsienko:2004bo}). They are $\mathbb{N}$ for $N=2$, three for $N$ odd, and two for $N>2$ even.

Based on the above, we propose below the finite temperature version of the $\textrm{SL}(N)$-invariant family of Schwarzian actions. To do so, we study in some detail the appearance of the Schwarzian theory in the $N=2$ case and then propose a generalization for $N>2$.

In the $\textrm{SL}(2)$ case, equation \eqref{eq:genHills} is identical to Hill's equation \eqref{eq:hills} with $\psi=\psi_{1,2}$. Suppose the parametization $\tau$ is such that $\cL$ is constant. The anti-periodic boundary conditions \eqref{eq:holo} force $\mathcal{L}$ to be $\mathcal{L}=1/4$. The two independent solutions read $\psi_1=\sqrt{2}\cos(\tau/2)$ and $\psi_2=\sqrt{2}\sin(\tau/2)$. For non-constant $\cL$, we build up a one-parameter family of solutions that have the same monodromy as the previous solution. By applying a diffeomorphism $\theta(\tau)$  on $\psi$ we find
 \eq{
 \hat{\psi}_1(\tau)=\sqrt{\frac{2}{\theta'(\tau)}}\cos\left(\frac{1}{2}\theta(\tau) \right)\,, \qquad \hat{\psi}_2(\tau)=\sqrt{\frac{2}{\theta'(\tau)}}\sin\left(\frac{1}{2}\theta(\tau)\right)\,.
 }{eq:Dorbit}
The corresponding $\cL$ associated to this orbit of solutions is given by 
\eq{
\cL= \left \{\cot\left(\frac{1}{2}\theta(\tau)\right);\tau \right\}\,.
}{eq:cL2}
Thus, using \eqref{eq:schwarzfamily} to define the action, we conclude that the Schwarzian theory is recovered. An important observation that will be crucial for the generalization is that the argument of $\cL$ in this approach is given by
\eq{
\hat{s}(\tau)=\frac{\hat{\psi}_1}{\hat{\psi}_2}=\cot\left(\frac{1}{2}\theta(\tau)\right)
}{eq:mapsl2}
which is precisely the map that relates the projective line to the circle $S^1$. Note that while $(\hat{\psi}_1,\hat{\psi}_2)\in\mathbb{R}^2$ is anti-periodic on a $2\pi$-period, the function $\hat{s}$ is periodic. This illustrates the relation between the monodromy \eqref{eq:Mg1} and the one associated to the lift \eqref{eq:Mg0}.

Motivated by the previous analysis, we consider
\eq{
\hat{s}_i: S^{N-1}\rightarrow \mathbb{RP}^{N-1}
}{eq:suerte}
which defines a projection of the coordinates $s_i$ into the unit sphere $S^{N-1}$ and satisfies the monodromy condition \eqref{eq:Mg1}. We propose that 
\begin{equation}
I[g]=\kappa \frac{k\bar{y}}{2\pi} \int\limits^{2\pi}_0\extd \tau\, \mathcal{L}[\hat{s}_1(\tau)\cdots\hat{s}_{N-1}(\tau)]\,,
\label{eq:schbeta}
\end{equation}
is the Schwarzian action at finite temperature $\beta=2\pi$ [with $\kappa$ defined in \eqref{eq:kappa}]. 

As an example, let us present the map \eqref{eq:suerte} associated to the $N=3$ case. This is given by the \emph{central} projection of $S^2$ on $\mathbb{RP}^2$
\eq{
\Big(\hat{s}_1,\hat{s}_2\Big)=\Big(\cot(\theta)\cos(\vp),\cot(\theta)\sin(\vp)\Big)\,.
}{eq:central}
The idea of this map is that we choose a point as the center of $S^2$ and a tangent plane to it representing $\mathbb{RP}^2$. Lines passing trough the center projects points $(\hat{s}_1,\hat{s}_2)$ on the two-sphere represented by $(\theta,\vp)$. Provided $\theta(\tau+2\pi)\sim \theta(\tau )+\pi$ and $\vp(\tau+2\pi)\sim\vp(\tau)+2\pi$, this map ensures that the monodromy condition is satisfied.

\subsection{One-loop contribution from higher-spin fields}

\label{se:one-loop}
We would like to explore the effect of considering higher rank groups in the one-loop contribution to the free energy.  Let us consider
\eq{
Z[\beta]= \int \extd\mu[g] \,e^{-I[g]}
}{z}
where $g\in{\rm SL}(N,\mathbb{R})$ and $\mu$ is a measure that we will leave unspecified for the moment. Action $I[g]$ in the partition function is given by
\eq{
I[g]= \frac{\kappa}{\sigma^2}\,\int\limits^{2\pi}_{0} \extd \vp \, \cL[\hat{s}_{1}(\vp),\cdots,\hat{s}_{N-1}(\vp)]\qquad\qquad \sigma^{-2}= \frac{k \bar{y} }{\beta}\,.
}{actlss}
where we have introduced the coordinate $\vp=\frac{2\pi}{\beta}\tau$ [and again $\kappa$ is defined in \eqref{eq:kappa}]. The one-loop contribution can be computed from the second variations associated to \eqref{actlss}. This can be expressed as
\eq{
\delta^2 I[g]= \,\int\limits^{2\pi}_{0} \extd \vp \, \tr\left[ ( \d_{\vp}\eps-[\d_{\vp}g g^{-1},\eps ] ) \d_{\vp}\eps \right]
}{variations}
where we defined $\delta g g^{-1}= \sigma \, \eps(\vp) $ and we have introduced the parameter $\sigma$ in the definition to keep the track of the perturbation expansion order. Evaluation in the gravitational sector amounts  to use the condition $\d_{\vp}g g^{-1}=-a^{\rm reg}_{\vp}$, where $a^{\rm reg}_{\vp}$ is the connection satisfying the regularity condition \eqref{eq:holo}. Thus, the quadratic fluctuations around the saddle are controlled by the second-order operator
\eq{
\Delta=-\d_{\vp}( \d_{\vp} +[a^{\rm reg}_{\vp},\cdot ] )\,.
}{op}
Operator $\Delta$ has $N^{2}-1$ zero modes corresponding to the ${\rm SL}(N,\mathbb{R})$  isometries of $a^{\rm reg}_{\vp}$. Summing over inequivalent configurations in \eqref{z} implies that we should consider this modes as gauge symmetries. Following \cite{Maldacena:2016hyu}, the path integral measure should be corrected with the introduction of the product $$\prod^{N^2-2}_{i=0} \delta(\eps^{(i)}(0))$$ which will remove the zero modes associated to $\Delta$. To evaluate the quadratic contribution to \eqref{z}, we must express the measure associated to $g$ in terms of $\eps$. This means that we need to trade every Fourier mode of $\delta g g^{-1}$  for a Fourier mode of $\eps$, except for the $N^2-1$ zero modes that have been fixed. Extending the argument of \cite{Stanford:2017thb} to the SL$(N)$ case, the result for this determinant is given by $\sigma^{1-N^2}$. In turn, the one-loop contribution to the free energy is
\eq{
F_{\tiny \textrm{1-loop}}= \frac{N^2-1}{2}\,T\,\log\left(k\bar{y} T \right)\,.
}{one-loop}

In the next section we discuss consequences for the entropy of the leading and subleading terms, i.e., the zero- and one-loop contributions to the partition function for the extensions of the Schwarzian action (Yang-Mills, higher spins). 

\section{Entropy}\label{se:entropy}

It is of interest to calculate the entropy associated with thermal states in BF-theories, as this corresponds to the black hole entropy in cases where a gravitational interpretation exists. We focus first on the leading, classical, contributions to the AdS$_2$ black hole entropy.

One can derive the entropy in a variety of ways. In the present context perhaps the simplest derivation is from evaluating the Euclidean on-shell action (see \cite{Grumiller:2007ju} and references therein), multiplying by temperature to get free energy 
\eq{
F(T) = T\,\big(I_0+I_B\big)_{\textrm{\tiny EOM}} = T\,I_B\big|_{\textrm{\tiny EOM}}
}{eq:syk58}
and then taking the $T$-derivative to get entropy [with $I_B$ determined from \eqref{act2} together with a suitable integrability condition \eqref{act3}].
\eq{
S = -\frac{\extd F}{\extd T} = - I_B\big|_{\textrm{\tiny EOM}} - T\,\frac{\extd I_B}{\extd T}\big|_{\textrm{\tiny EOM}}
}{eq:syk57}

It is obvious, though still remarkable and different from the generic situation in higher dimensions, that the entropy is determined entirely by boundary data, namely the on-shell value of the Schwarzian action \eqref{act99} together with the periodicity condition \eqref{eq:syk59} (or corresponding generalizations thereof). By contrast, in, say, three-dimensional (higher spin) gravity entropy is obtained from a boundary term at the horizon \cite{Banados:1993qp, Hawking:1994ii, Teitelboim:1994az, Banados:2012ue, deBoer:2013gz, Compere:2013nba, Bunster:2014mua, Grumiller:2017otl}, as anticipated on general grounds \cite{Wald:1993nt, Iyer:1994ys}. The reason for this difference is essentially captured by Fig.~\ref{fig:1}; after imposing regularity there is only a contractible cycle, while in three or higher dimensions non-contractible cycles remain in general and are associated with properties of the horizon rather than the asymptotic region.

In the spin-2 case we recover in this way from the on-shell action \eqref{eq:syk10} the result of \cite{Grumiller:2015vaa, Maldacena:2016upp, Grumiller:2017qao},\footnote{%
To compare with \cite{Maldacena:2016upp} we need to identify $\bar \phi_r$ in their (3.18) with our $\bar y$; to compare with \cite{Grumiller:2015vaa, Grumiller:2017qao} we need to rescale the coupling constant $k_{\textrm{\tiny here}}=2k_{\textrm{\tiny there}}$.
}
\eq{
S_{\textrm{\tiny JT}} = \frac{k}{2\pi\bar y}\,\frac{\extd C}{\extd T} = k\pi\bar y\,T
}{eq:syk60}
where we have used the regularity condition \eqref{eq:syk59} (setting $n=1$) together with the relation \eqref{act10} between Casimir $C$ and mass function $\cL$. Note that the result for entropy \eqref{eq:syk60} is compatible with the third law of thermodynamics and shows the same temperature dependence as a Fermi-liquid (or -gas) with Sommerfeld constant\footnote{%
The Sommerfeld constant is the ratio of specific heat and temperature in the limit $T\to 0$, which in the Fermi-liquid case reduces to the coefficient linear in $T$ in the small-$T$ expansion of the entropy.} 
given by $\gamma=k\pi\bar y$. 

The gravity result \eqref{eq:syk60} for the entropy coincides qualitatively with the field theory result derived in \cite{Maldacena:2016hyu}, see their (G.241): the first term in their expansion is temperature-independent and captures the zero-temperature entropy $S_0$ that is not modeled by JT. The second term in their expansion
\eq{
S-S_0 = 2a_3 \frac{NT}{J} \stackrel{?}{=} S_{\textrm{\tiny JT}}
}{eq:G241}
should then correspond to the entropy \eqref{eq:syk60}. Using the holographic dictionary \eqref{eq:syk11} shows that indeed these two expressions coincide for any $N, J, T$ (subject to $N\gg 1$ and $J\gg T$) for some value of the numerical coefficient $a_3$ that is independent from $N,J,T$.  

For the Yang--Mills generalization discussed in section \ref{se:4} we find an additional contribution to entropy,
\eq{
S_{\textrm{\tiny JT,\,YM}} = k\pi\bar y\, T\,\big(n^2-4m^2\big)
}{eq:syk56}
where we used the relation \eqref{u8} between Casimir and state-dependent functions and the relations \eqref{act9.2}, \eqref{eq:syk80} between state-dependent functions and temperature. Thus, the only solution of the holonomy condition [on a single cover of AdS$_2$, i.e., for $n = 1$] compatible with a positive entropy for arbitrary temperatures is given by $m = 0$, which implies $\cP = 0$ and hence recovers the JT result for the entropy \eqref{eq:syk60}. 

By virtue of \eqref{eq:syk21} the free energy for the spin-3 generalization discussed in section \ref{se:5} is given by [$f_0^{(2,\,3)}$ denote zero modes of the functions $f^{(2,\,3)}$]
\eq{
F_{(3)} = -\frac{k}{2\pi}\,\big(f^{(2)}_0C_2 + f^{(3)}_0C_3\big)
}{eq:syk65}
and leads to the corresponding entropy
\eq{
S_{(3)} = \frac{k}{2\pi}\,\Big(f_0^{(2)}\,\frac{\extd C_2}{\extd T} + f_0^{(3)}\,\frac{\extd C_3}{\extd T}\Big)\,.
}{eq:syk66}

The scaling properties of the differential equation \eqref{eq:syk24.2} with Euclidean time $\tau$ imply that $\cL$ (and hence $C_2$) should scale like $T^2$ and that $\cW$ (and hence $C_3$) should scale like $T^3$. With these scalings the entropy \eqref{eq:syk66} can be written as
\eq{
S_{(3)} = \frac{k}{2\pi}\,\Big(f_0^{(2)}\,\frac{2 C_2}{T} + f_0^{(3)}\,\frac{3 C_3}{T}\Big)\,.
}{eq:syk67}

More generally, the free energy for spin-$N$ theories should be given by a sum of Casimirs
\eq{
F_{(N)} = -\frac{k}{2\pi}\,\sum_{s=2}^N f_0^{(s)}C_s
}{eq:keep1}
where $f^{(s)}_0$ denotes the zero mode of the function $f^{(s)}$ associated with a field of spin $s$.
Since we expect the scaling behavior $C_s\propto T^s$ the entropy then would be
\eq{
S_{(N)} = \frac{k}{2\pi}\,\sum_{s=2}^N f_0^{(s)} \frac{s\, C_s}{T} \sim \sum_{s=2}^N \hat f^{(s)} T^{s-1} \,.
}{eq:keep2}

In all cases above at low temperatures the entropy is dominated by the spin-2 contribution (as long as $f_0^{(2)}\neq 0$), which scales linearly in $T$. In particular, the JT-result for entropy \eqref{eq:syk60} receives modifications from higher spin fields only at higher temperatures.

We consider now the 1-loop contribution to the entropy. The general expression \eqref{eq:syk57} together with the classical \eqref{eq:keep2} and 1-loop results \eqref{one-loop} yields
\eq{
S_{\textrm{\tiny 1-loop}} = S_{(N)} - \frac{N^2-1}{2}\,\ln S_{(N)} + {\cal O}(1)\,.
}{eq:syk99}
For the SL$(2)$ case the famous factor $-3/2$ (see e.g.~\cite{Sen:2012dw} and refs.~therein) in front of the log-term is recovered, while for general SL$(N)$ this factor is instead $-(N^2-1)/2$. The result \eqref{eq:syk99} implies that the dominant contribution from higher spin fields in the small temperature limit actually may come from the 1-loop contribution to the entropy, as the classical contribution is suppressed by $T^{s-1}$.

\section{Summary and (generalized) SYK/JT holographic dictionary}\label{se:7}

We summarize our main results. In the BF-formulation of the (generalized) Jackiw--Teitelboim model reviewed in section \ref{se:2} we saw that the observables where either Wilson loops around the temperature cycle \eqref{eq:angelinajolie} or (on-shell conserved) Casimirs \eqref{casn}. The action principle is well-defined upon imposing an integrability condition \eqref{act3}, which for JT (discussed in section \ref{se:3}) is given by \eqref{act4}, thereby recovering previous results \cite{Grumiller:2017qao}. The on-shell action (which then determines free energy and entropy) is proportional to inverse temperature, Casimir and inversely proportional to a zero mode associated with the dilaton at the asymptotic AdS$_2$ boundary \eqref{eq:syk10}. This boundary action can be rewritten as Schwarzian action \eqref{act99}, as expected \cite{Maldacena:2016upp}. Regularity of the Wilson-loops requires a quantization \eqref{eq:syk59} in units of temperature squared of the free parameter in the BF connection \eqref{FGV}.

We then extended these results in section \ref{se:4} to the JT model coupled to Yang--Mills fields and obtained a generalized Schwarzian action \eqref{act9} together with new quantization conditions \eqref{act9.2}, \eqref{eq:syk80}.

In section \ref{se:5} we provided steps towards higher spin SYK, working exclusively on the gravity side. While we were again able to find a well-defined action principle, this did not provide us with enough information to derive the higher spin analogue of the Schwarzian action. Thus, in section \ref{se:6} we followed a different approach, studying the symplectic structure of (generalized) JT. The boundary action \eqref{qua} is the action of a particle on a group manifold (e.g.~SL$(3,\,\mathbb{R})$ for spin-3 gravity), and is subject to two additional constraints, namely asymptotic AdS$_2$ behavior (with fluctuating dilaton \cite{Grumiller:2013swa, Grumiller:2015vaa}) and regularity of the Wilson loops. Solving the former yields generalized Schwarzian actions, which we have displayed explicitly for the spin-2 and spin-3 case \eqref{eq:schwarziansl2}. Solving the latter \eqref{eq:holo} leads to specific boundary conditions for the fields appearing in the Schwarzian action. As an extension of the arguments of \cite{Stanford:2017thb}, in section \ref{se:one-loop} we determined the contribution of higher spin fields to the one-loop partition function. 

In section \ref{se:entropy} we discussed thermodynamical aspects, in particular the entropy, which is determined directly from the on-shell action \eqref{eq:syk57}. The contribution of fields of arbitrary spin $s$ to the entropy scales like $T^{s-1}$ \eqref{eq:keep2}, so that at low temperatures the dominant non-constant behavior comes from the spin-2 field. For future reference we collect various entries of the SYK/JT holographic dictionary in table \ref{tab:1}. 
 \begin{table}
\begin{center}
\noindent \begin{tabular}{|l|l|}\hline
{\bf Gravity (JT and generalizations)} & {\bf Field theory (SYK and generalizations)} \\ \hline
$k\sim 1/G$ (grav.~coupling strength) & $\sim N$ (number of Majorana fermions) \\
$1/\bar y$ (zero mode in integrability condition) & $\sim J$ (strength of random coupling) \\
$\beta$ (length of Euclidean time cycle) & $1/T$ (inverse temperature) \\
$\ell_f\gg 1$ ($f$-cycle decompactifies) & $T\ll J$ (small temperature limit) \\
boundary action \eqref{act99} & Schwarzian action \eqref{eq:syk1} \\
Casimir function[s] \eqref{casn} & [generalized] Schwarzian \eqref{act8} [\eqref{u8}, \eqref{eq:schwarziansl2}] \\
off-shell asymptotic symmetries & reparametrizations of temperature cycle \\
absence of holonomies & symmetry breaking to SL$(2,\mathbb{R})$ \\
allowed values of black hole mass function & elements of orbits of Diff$(S^1)/$SL$(2,\mathbb{R})$ \\
$\cL$ (black hole mass) & $\sim T^2$ (temperature squared)  \\ 
$C$ (quadratic Casimir, on-shell) & $\sim T^2/J^2$ (dimensionless temperature squared)\\ 
$S_{\textrm{\tiny black}} = k\pi\bar y\, T$ (black hole entropy) & $S_{\textrm{\tiny Schwarz}}\sim N\, T\,/J$ (Schwarzian entropy) \\
$k\pi\bar y$ (holographic Sommerfeld constant) & $\sim N/J$ (Sommerfeld constant) \\
boundary gravitons & Goldstone bosons \\
Yang--Mills fields & additional conserved charges \\
higher spin fields & contribution to log-corrections of entropy \\  \hline
\end{tabular}
\smallskip
{\footnotesize The similarity signs refer to ${\cal O}(1)$ factors that are independent from $N, J$ and $T$.} 
\end{center}
\caption{Generalized SYK/JT correspondence}
\label{tab:1}
\end{table}

Note in particular the entropy relation \cite{Maldacena:2016upp}
\eq{
S_{\textrm{\tiny black}} \sim S_{\textrm{\tiny Schwarz}}
}{eq:S}
where $S_{\textrm{\tiny black}}$ is the JT black hole entropy and $S_{\textrm{\tiny Schwarz}}$ is the field theory entropy in the small $T$ limit with the $T=0$ result subtracted. As shown in section \ref{se:entropy} the same relation remains true at small temperatures after including Yang--Mills or higher spin fields. It is therefore not clear if there is a field theory generalization of SYK accessible in the regime $T\ll J$ that is sensitive to higher spin fields. However, even at small temperatures higher-spin fields in principle are detectable semi-classically through a change of the numerical coefficient in the log-corrections to the entropy \eqref{eq:syk99}. It could be thus very interesting on the field theory side to generate SYK-like models where this coefficient in the log-corrections to the entropy can be tuned to $-(N^2-1)/2$, where $N$ is some integer, in order to mimick the behavior of spin-$N$ theories in AdS$_2$.

Finally, we mention some possible further developments. Besides BF theories also Poisson-sigma models where the Poisson-tensor has one constant entry have an interpretation as two-dimensional dilaton gravity \cite{Verlinde:1991rf, Cangemi:1992bj}, namely the conformally transformed string black hole in two dimensions \cite{Mandal:1991tz, Elitzur:1991cb, Witten:1991yr} (see also \cite{Callan:1992rs}), the corresponding algebra being centrally extended Poincar\'e (also known as Maxwell algebra, see e.g.~\cite{Bonanos:2009wy}). It should be straightforward and might be interesting to generalize our analysis --- and the corresponding quantum mechanical side of the story --- to this case. One could also consider higher spin theories using non-principal embeddings of sl$(2)$ into sl$(N)$ or into other gauge algebras.

Of course, it would even be better to generalize our results to arbitrary Poisson-sigma models (at least those with an asymptotically AdS$_2$ interpretation), which requires to deal with non-linear algebras and their associated groups. If successful, a large class of models can be described, some of which emerge from dimensional reduction of gravity in arbitrary dimensions, see e.g.~the list of dilaton gravity models in table 1 of \cite{Grumiller:2006rc}. Furthermore, this may provide the path to establish SYK/JT-like correspondences describing on the gravity side for instance the s-wave sector of Einstein gravity, i.e., to find a field theory dual for the Schwarzschild-AdS black hole.

\acknowledgments

We thank Robert McNees, Carlos Valc\'arcel and Dima Vassilevich for collaboration on holographic (and other) aspects of dilaton gravity in two dimensions. We thank Marcela C\'ardenas, Oscar Fuentealba, Joaquim Gomis, Wout Merbis, Friedrich Sch\"oller and Ricardo Troncoso for useful discussions. 

This work was supported by the Austrian Science Fund (FWF), projects P~27182-N27 and P~28751-N27. DG thanks the Centro  de  Estudios  Cient\'ificos (CECs) for hospitality during the final stage of this work. CECs  is  funded  by  the  Chilean  Government  through the  Centers  of  Excellence  Base  Financing  Program  of Conicyt.

\paragraph{Note added.} While finishing this paper the work \cite{Mertens:2018fds} appeared, which focuses mostly on the relation between three- and two-dimensional gravity, a topic that we do not consider. However, there is overlap in one aspect and our respective results agree with each other. Namely the bosonic sector of the $N=2$ super-Schwarzian discussed in \cite{Mertens:2018fds} is related to our Yang--Mills generalization for the special case where the gauge group is abelian.

\appendix

\section{Spin-3 Casimirs}\label{app:C}

For the $SL(3,\mathbb{R})$ case we have a quadratic Casimir $C_2$ \eqref{eq:syk19}, which expressed in terms of $y,z,\cL,\cW$ reads
\begin{multline}
C_2=-\frac{64}{3}\cL^2 z^2 + \frac{20}{3}\cL\big((z')^2-2 z z''\big)-\frac{20}{3} \cL' z z'  - \frac{8}{3} \cL'' z^2 + 4\cL y^2+ 12 \cW z y' \\ -\frac{1}{3} (z'')^2+\frac{2}{3}z'z''' -\frac{2}{3}z z^{(4)} +2yy''- (y')^2
\label{eq:C2}
\end{multline}
and a cubic Casimir $C_3$ \eqref{eq:syk19}, the explicit form of which is given by
\begin{multline}
C_3=-\frac{1024}{27}\cL^3 z^3 +\frac{64}{9}\cL^2\big(3z (z')^2 -5z^2 z''\big)  -\frac{32}{3} \cL \cL' z^2 z' +\frac{32}{9} \big(\cL'^2 -2\cL \cL'' -9 \cW^2\big) z^3-\frac{64}{3} \cL^2 y^2 z\\
+32 \cL \cW y z^2 +\frac{4}{9} \cL \big(7(z')^2 z'' - 14 z (z'')^2 + 8 z z' z^{(3)}  - 4z^2 z^{(4)} \big) +\frac{2}{9} \cL' \big(21(z')^3 - 32 z z' z'' + 8z^2 z^{(3)} \big) \\
+\frac49\cL''  z \big(3(z')^2 - 4z z'' - 3 y^2\big)  + \frac{4}{3}\cL \big(8 y y' z' - 2 z (y')^2 -5 y^2 z''- 8 y z y''\big)  + \frac23 \cL' y \big(4 z y' -7y z'\big) \\
+2\cW \big(-y^3 - 4 z y' z'+ y (z')^2 + 8 z^2 y''\big) +\frac{2}{9} z (z^{(3)})^2 -\frac{4}{9} z  z'' z^{(4)} +\frac{2}{27}(z'')^3+\frac{1}{3}  (z')^2 z^{(4)}-\frac{2}{9}  z' z'' z^{(3)} \\
 -\frac{1}{3} y^2 z^{(4)} + \frac{2}{3} y y'z^{(3)} - \frac{2}{3} y y'' z'' -2 z y''^2 - \frac{2}{3} (y')^2 z'' + 2 y' y'' z'\,.
\end{multline}
For the constant representative $\cL=1/4$, $\cW=0$ these expressions simplify.
\eq{
C_2\big|_{\cL=1/4,\cW=0}=-\frac{4}{3}\, z^2 - \frac{10}{3}\, z z'' -\frac{2}{3}\,z z^{(4)} + \frac{5}{3}\,(z')^2 + \frac{2}{3}\,z'z''' -\frac{1}{3}\, (z'')^2 + y^2 +2yy''- (y')^2
}{eq:c2w}
\begin{multline}
C_3\big|_{\cL=1/4,\cW=0} = -\frac{16}{27}\, z^3 - \frac49\,z^2\,\big(5z'' + z^{(4)}\big) + \frac29\,z\,\big(6(z')^2 - 7(z'')^2 + 4 z' z^{(3)} - 2 z'' z^{(4)} + (z^{(3)})^2\big)\\
+ \frac19\,(z')^2 \,\big(7 z'' + 3z^{(4)}\big)  - \frac{2}{9}\, z' z'' z^{(3)} + \frac{2}{27}\,(z'')^3  - \frac13\,y^2\,\big(4z + 5z'' + z^{(4)}\big) \\
+ \frac23\,y\,\big(4y' z' + y' z^{(3)} - 4 y'' z  - y'' z''\big) - \frac{2}{3} \, (y')^2\,\big(z + z''\big) + 2 y' y'' z' - 2 (y'')^2 z   \,.
\end{multline}


\end{document}